%% file: main.tex
\documentclass[12pt,a4paper,fleqn]{article}
\usepackage[utf8]{inputenc}
\usepackage{amsmath}
\usepackage{breqn}
\usepackage{authblk}
\usepackage{amsfonts}
\usepackage{amssymb}
\usepackage{amsthm}
\usepackage{graphicx}
\usepackage{multirow}
\usepackage{import}
\usepackage{caption}
\usepackage{subcaption}
\usepackage{booktabs}
\usepackage{tabularx}
\usepackage[hidelinks]{hyperref}
\usepackage{cleveref}
\usepackage{tikz}
\newtheorem{theorem}{Theorem}[section]

\newtheorem{lemma}[theorem]{Lemma}
\newtheorem{definition}{Definition}[section]
\usepackage{xcolor, soul}
\hypersetup{
    colorlinks=false, 
    linktoc=all 
}
\usepackage[skip=2pt,font=scriptsize]{caption}
\usepackage{setspace}
\usepackage[vlined, ruled]{algorithm2e}
\DontPrintSemicolon
\SetAlgoLined
\usepackage{float}
\usepackage[flushleft]{threeparttable}
\usepackage[left=3cm,right=3cm,top=2cm,bottom=2cm]{geometry}
\usepackage[nodoi]{apacite}
\usepackage{natbib}
\usepackage{standalone}
\usepackage{subfiles}

\title{On non-negative auto-correlated integer demand processes}
\author[a,b,*]{Lotte van Hezewijk}
\author[a]{Nico P. Dellaert}
\author[a]{Willem L. van Jaarsveld}
\affil[a]{Industrial Engineering \& Innovation Sciences \\Eindhoven University of Technology \\P.O. Box 513, Eindhoven, 5600 MB, The Netherlands}
\affil[b]{ORTEC B.V., Houtsingel 5, Zoetermeer, 2719 EA, The Netherlands}
\affil[*]{Corresponding author: l.v.hezewijk@tue.nl}
\date{}
\begin{document}
\maketitle
\begin{abstract}
 Methods to generate realistic non-stationary demand scenarios are a key component for analyzing and optimizing decision policies in supply chains. Typical forecasting techniques recommended in standard inventory control textbooks consist of some form of simple exponential smoothing (SES) for both the estimates for the mean and standard deviation. We study demand generating processes (DGPs) that yield non-stationary demand scenarios, and that are \emph{consistent} with SES, meaning that SES yields unbiased estimates when applied to the generated demand scenarios. As demand in typical practical settings is discrete and non-negative, we study consistent DGPs on the non-negative integers. We derive conditions under which the existence of such DGPs can be guaranteed, and propose a specific DGP that yields autocorrelated, discrete demands when these conditions are satisfied.

Our subsequent simulation study gains further insights into the proposed DGP. It demonstrates that from a given initial forecast, our DGPs yields a diverse set of demand scenarios with a wide range of properties. To show the applicability of the DGP, we apply it to generate demand in a standard inventory problem with full backlogging and a positive lead time. We find that appropriate dynamic base-stock levels can be obtained using a new and relatively simple algorithm, and we demonstrate that this algorithm outperforms relevant benchmarks. 
\end{abstract}
\section{Introduction}\label{sec:introduction}
\subfile{input/1-Introduction.tex}

\section{Literature review}\label{sec:literature}
\subfile{input/2-LiteratureReview.tex}

\section{The demand generating process}\label{sec:demandmodel}
\subfile{input/3-DemandModel.tex}
\section{Application in inventory control}\label{sec:problemdescription}
\subfile{input/4-ProblemDescriptionInventory.tex}

\section{Conclusion}\label{sec:conclusion}
\subfile{input/6-Conclusion}


\bibliographystyle{apacite}
\bibliography{bib}
\end{document}

%% file: input/1-Introduction.tex
Companies throughout the world face many forms of uncertainty that impact operational supply chain processes. One of these sources of uncertainty is the demand for their products. In order to evaluate a variety of decision policies for supply chain operations planning problems like inventory control or capacity planning, companies need realistic demand scenarios. Realistic demand scenarios are essential for evaluating policies or strategies in dynamic supply chain problems, and they can also be used in optimization methods such as deep reinforcement learning (DRL; \citealt{boute2022deep}), (multi-stage) stochastic programming \citep{kataoka1963stochastic}, robust optimization \citep{Bertsimas2019} and bootstrapping \citep{Snyder2002,Boylan2022,Goltsos2022}. As demand is typically non-stationary, it does not suffice to estimate a distribution based on some demand observations, and to use that fitted distribution to generate a demand realization for each period independently. Instead we require a way of simulating scenarios that evolve over time. This need is recognized widely, and ARIMA models \citep{hyndman2018forecasting} such as the one used by \citet{Graves1999} and \citet{Gilbert2005} are extensively used for representing non-stationary demand processes in inventory control research. 

\begin{figure}[h] 
\centering
\includegraphics[width=\textwidth]{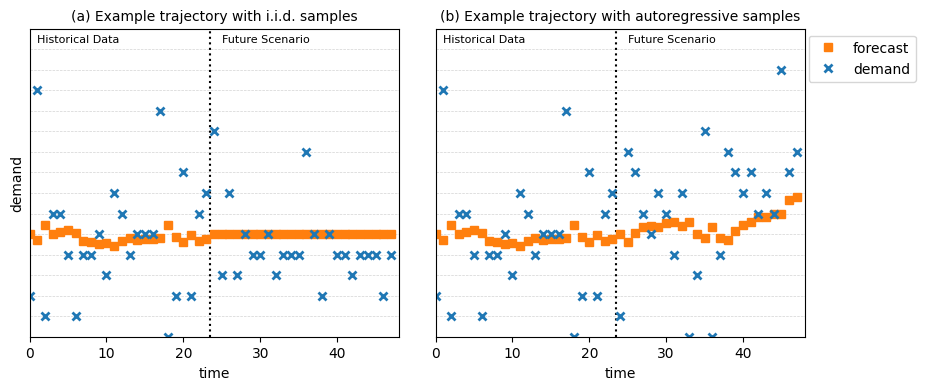}
\caption{Example of demand trajectory range when drawing i.i.d. samples (a) vs. autoregressive samples (b).}
\label{fig:intuition}
\end{figure}

To see the importance of non-stationarity in demand processes, consider for instance the case where one wants to use DRL or some type of simulation-based optimization to find good inventory replenishment policies. In order to train the algorithm or estimate the impact of policies, samples of demand trajectories are required. We have some data about historic demand ($d_t$) and the forecasts for those periods ($f_t$). In papers focusing on applying DRL in supply chain problems, some demand distribution is assumed, and demand trajectories are generated by sampling from this stationary distribution \citep{boute2022deep}. So in that case, the historical demand data could be used to derive some parametric demand distribution using the first- and second-order statistics. Another approach (see for example \citealt{Altendorfer2016}) is to use also the forecast history to fit a distribution for the forecast error, to sample from the stationary forecast error distribution and to multiply this with the latest forecast. Both of these approaches will result in a stable demand trajectory range over time, as illustrated in Figure \ref{fig:intuition} (a). 

However, knowing that companies employ forecasts and update those over time based on more recent observations, it is inconsistent to draw demand trajectory samples from stationary and i.i.d. distributions. Instead, to accurately represent reality, demand scenarios should be \emph{autoregressive}, which would result in a different demand trajectory range, as illustrated in Figure \ref{fig:intuition} (b), which also reflects increasing demand uncertainty the longer we look into the future. With positive autocorrelation, if demand in early periods turns out higher than initially expected, then demand in later periods is likely to also be higher than initially anticipated. With negative autocorrelation, high demand in certain periods will be followed by periods with lower demand. That real-life demand processes are autoregressive is well-known, leading to the widespread use of ARIMA models for representing demand processes. 


The prevalence of autoregressive demand processes is also recognized by seminal books on inventory control \citep{Silver2017,Axsater2015}, which recommend forecasting via exponential smoothing as a suitable technique for most products. With this method, positive autocorrelation is the underlying assumption. Standard simple exponential smoothing yields point forecasts. While well-studied in forecasting literature, point forecasts are insufficient for properly balancing underage and overage costs in inventory control settings \citep{Axsater2015}. For that latter purpose, estimating the standard deviation $\sigma$ as well as the mean $\mu$ of demand is crucial, and standard inventory textbooks \citep{Silver2017,Axsater2015} recommend to estimate the second moment of demand by smoothing the forecast error over time, which yields an approach that we shall refer to as simple exponential smoothing for mean and standard deviation (SES). The forecasting literature has traditionally focused on point estimates, but the idea of forecasting variance has been proposed there as well, see e.g. \citet{Gelper2010}.

The broad application of SES in inventory practice and recommendations by standard textbooks \citep{Silver2017,Axsater2015} motivate us to study demand processes that are consistent with SES. Intuitively, this consistency entails that SES is a suitable forecasting process for the demand scenarios resulting from the demand generating process (DGP). That is, SES applied to the first $t$ data points $\{d_1,...,d_t\}$ yields unbiased estimates of the mean and variance of the demand in period $t+1$. DGPs that are consistent with SES yield demand scenarios that are autoregressive. See also \cite{Graves1999}, who shows that standard exponential smoothing is consistent with an ARIMA $(0,1,1)$ process.

For demand models to be practically useful, they would benefit from some properties in addition to being autoregressive. In particular, \citet{Kolassa2016}, \citet{Kaya2018} and \citet{Rostami-Tabar2023} emphasize the need for discrete, non-negative demand distributions in supply chain management, especially for low-volume demand, for example for safety stock setting, but also scenario analysis or other forecast-driven decision processes. Typical ARIMA models fall short in this regard: they tend to feature time series with continuous and negative values. 
More importantly, the error terms which cause the `shocks' in the demand process are generally assumed to be independent, identically distributed variables \citep{Graves1999,Gilbert2005}, which is contrary to the idea of updating estimates of $\sigma$ over time. 

When the demand can take on any real value, specifying a demand process that is consistent with any SES forecasting procedure is relatively straightforward, since for any combination of $\mu$ and $\sigma$ there exists a corresponding continuous distribution (for example a normal distribution). However, not all combinations of mean and variance have a corresponding distribution on the non-negative integers \citep[see][]{Adan1995}. A natural question that arises is whether there exist demand processes that are non-negative, discrete \emph{and} consistent with SES forecasting. 

The main contribution of this paper is providing an answer to this question. In particular, we derive the conditions needed for existence of consistent demand processes on the non-negative integers. As long as we adhere to the conditions for $\mu$ and $\sigma$ outlined by \cite{Adan1995}, and use identical smoothing parameters ($\alpha = \beta$) for updating the mean and variance, we can generate demand processes that are non-negative, discrete \emph{and} consistent with SES. When the smoothing parameters are not identical, situations can arise where the $\mu$ and $\sigma$ no longer satisfy the conditions, meaning that their distribution on the  non-negative integer domain does not exist \citep{Adan1995}.

When the conditions for $\mu$, $\sigma$ and the smoothing parameters are satisfied, the developments in this paper yield an intuitive and simple DGP that (1) is consistent with simple exponential smoothing as proposed by the standard textbooks on inventory control \citep{Axsater2015, Silver2017}, (2) takes on values on the non-negative integers, and (3) considers updating the estimate of the demand variance rather than assuming known and exogenous variance (4) yields demand scenarios that are autoregressive, covering non-stationary developments of demand, such as exploding or vanishing demand trajectories, and (5) covers a variety of demand profiles (e.g. erratic or intermittent) as often seen in practice. The DGP can generate demand scenarios for any feasible initial estimate of the mean and standard deviation of demand, and we view it as a key contribution of this paper.

To illustrate the relevance of the DGP, we demonstrate how these demand scenarios can be used in one specific type of supply chain operations planning problem: spare parts inventory control. Spare parts demand is known for its intermittency: due to many zero values in the times series, and a generally low volume of the demand, average demand is often close to zero and has a high coefficient of variation \citep{Amniattalab2023}. Continuous demand distributions such as the approach proposed in \citet{Graves1999} may be less appropriate for modeling spare parts demand, and our proposed DGP is expected to bring benefits under these circumstances.  To validate this, we compare the approach proposed by \citet{Graves1999} for determining base-stock levels in case of non-stationary ARIMA demand, with an alternative solution method that utilizes the knowledge about the DGP to determine base-stock levels that are better suited for the low-volume demand processes that often arise for spare parts demand.


The remainder of the paper is organized as follows: \Cref{sec:literature} provides an overview of the treatment of non-stationary demand processes for supply chain operations planning problems in literature. The DGP is proposed and analyzed in \Cref{sec:demandmodel}, where we also derive conditions under which the existence of the DGP can be guaranteed. In \Cref{sec:problemdescription} we describe the application of our DGP in the context of a simple inventory control problem, propose the solution method for finding base-stock levels, and examine the performance of this solution method using numerical experiments. We conclude the paper in \Cref{sec:conclusion}.

%% file: input/2-LiteratureReview.tex
 
The literature concerning demand processes can be roughly grouped into (1) stationary and independent and identically distributed (i.i.d.) demand and (2) non-stationary demand. Although i.i.d. stationary demand processes are popular because of their analytical tractability, they are often a poor reflection of reality. In this paper, we therefore focus on the body of literature for the latter category. Within non-stationary demand, we recognize two different streams of literature: (1) independent non-stationary demand and (2) autoregressive demand. Within the autoregressive demand stream, there are Markov-modulated and ARIMA-type demand processes. The contribution of this paper falls in the latter category. 

While we aim to focus mostly on non-stationary demand, we call some attention to studies that investigate how best to forecast lead time demand variance for stationary demand processes. For example in inventory control problems, a common way of determining safety stock levels in practice is to assume that the demand follows a normal distribution, and then to multiply the lead time demand standard deviation with some safety factor to obtain a certain service level \citep{Prak2017}. Note that for other types of distributions, it is possible to use the cumulative distribution function to find the value that corresponds to achieving a target service level. However, it is notoriously difficult to find the appropriate standard deviation of lead-time demand, even if the demand is stationary. The standard way of taking $\sigma \sqrt{L}$ as the standard deviation of lead-time demand ignores correlation of forecast errors and leads to an underestimation of the variance and thus safety stocks. \citet{Prak2017} and \citet{Janssen2009} demonstrate this and propose ways of correcting for this underestimation in stationary demand processes, in order to achieve the service level target.

In many papers considering non-stationary demand in operational planning problems, it is assumed that the demand distribution and its parameters are known, and independently varying over the periods in the time horizon. \citet{Xiang2018} compute near-optimal parameters of an ($s_t, S_t$) lot sizing policy using mixed integer linear programming. The authors assume that the demand in a period follows a known probability density function. While they illustrate their results using a normal distribution, they argue that their modeling strategy is distribution independent. The coefficient of variation remains fixed in every time period, and the mean in each time period is given by some non-stationary pattern (e.g. a product life cycle pattern or sinusoidal oscillations). Furthermore, the distributions in each period are independent. Knowing that the majority of companies use time series forecasting methods to predict the demand over time, it is unlikely that demand in different periods is independent. There are a number of studies that employ similar ways of modeling the non-stationary demand when aiming to optimize lot sizing policy parameters, and the reader is referred to \citet{Ma2022} or \citet{Visentin2021} for an extensive overview of those papers. 

Another way of modeling the fluctuating demand environment was proposed by \citet{Song1993}, where they consider that there are some identifiable factors (e.g. economic conditions, innovation) that determine the demand environment. This environment is modeled as a continuous-time Markov chain, and if the environment is in state $i$, demand follows a Poisson process with rate $\lambda_i$. This can be seen as an autoregressive demand process. With this generalization of standard demand processes, the authors provide qualitative descriptions of the forms of optimal inventory policies. Two algorithms are presented for computing the parameters of these policies. Other authors extend these Markov-modulated demand processes by investigating the impact of having partial observability of the demand \citep{Bayraktar2010}, or demand correlations \citep{Hu2016} on the structure of the optimal policy. They find that in such cases, the optimal policy is a state-dependent ($s, S$) policy, which is promising as these are widely used in practice. However, finding the state-dependent parameters is not trivial, and analytically intractable.

Several authors have also addressed the issue of estimating the lead-time demand variance in non-stationary ARIMA-type processes. \citet{Graves1999} models the demand as a (0,1,1) IMA process, where every period there is a shift in the mean of the demand process that is proportional (with factor $\alpha$) to the size of the shock which is normally distributed with mean $0$ and variance $\sigma^2$. Using this model for demand, the author provides a correction factor for the safety stock that is a function of the lead-time ($L$) and $\alpha$. \citet{Graves1999} concludes that `we require dramatically more safety stock when demand is non-stationary, in comparison with the textbook case of stationary demand' (p. 54). \citet{Strijbosch2011} use a similar model for the demand (the (0,1,1) ARIMA model), but they ensure that demand cannot be negative by truncating the mean. They conclude that most benefits in stock control performance can be attributed to having an appropriate demand variance updating procedure, rather than the choice of forecasting method or forecast parameter values. \citet{Prak2019} provide a framework on how to incorporate uncertainty about future demand into inventory models, using trended and random walk models for demand, that is normally distributed. \cite{ZiedBabai2022} explore three strategies to estimate variance of lead-time demand, and derive analytical results for an ARMA(1,1) demand process. Some research focuses on the need of predicting full lead-time demand distributions in supply chain operations planning. \citet{Cao2019} propose a neural network model to forecast quantiles of a non-negative non-stationary autoregressive demand process. 

A key feature of many autoregressive demand processes in literature is that they consider the \emph{variance} of the underlying demand process to be exogenous and known. In practical settings, the variance of the demand is unknown and rather forecasted and updated based on demand observations, in line with the recommendations by seminal inventory control books \citep{Axsater2015,Silver2017}. Another feature of many studied autoregressive non-stationary models is that they are all considering real-valued demand, as opposed to discrete, integer-valued demand that is typically seen in practice. We contribute a DGP that addresses two gaps in literature: (1) the need for a practical discrete, non-negative, and non-stationary autoregressive stochastic process that (2) includes updating the forecast of the expectation as well as the demand variance as recommended by \citet{Axsater2015} and \citet{Silver2017}. We contribute to literature by specifying a new demand process on the non-negative integers in which $\sigma$ is a forecast rather than an exogenously given number. In particular, $\sigma$ in our model is treated in the same way as the mean demand $\mu$: as an unknown parameter for which an initial estimate is available that will be updated in a manner consistent with simple exponential smoothing \citep{Axsater2015, Silver2017}. Most importantly, we derive the conditions required for the existence of such a consistent DGP on the non-negative integers. We show several desirable properties of the model, and illustrate a potential way in which it can be used to determine inventory control policies.

%% file: input/3-DemandModel.tex
Our proposed DGP is closely related to the exponential smoothing forecasting approach that shall be formalized in Section~\ref{sec:expsmooth}. The key ideas underlying the DGP will be introduced in Section~\ref{subsec:demand-updating}, and some key analytical results are provided in Section~\ref{sec:feasibility}. In Section~\ref{subsec:longtermestimates} we reflect on the long term demand expectations of the proposed DGP. 

\subsection{Simple exponential smoothing}\label{sec:expsmooth}
Simple exponential smoothing (SES) is a suitable forecasting approach when there is no clear trend or seasonality in demand. The DGP proposed in this paper is closely related to the SES forecasting method. To formally define SES, let $d_1,d_2,\ldots$ denote the demand time series, let $\alpha,\beta\in [0,1]$ denote two smoothing constants. Also, let $f_{t}$ and $MSE_{t}$ denote respectively our forecast of the mean demand in period $t$ and of the associated mean square error, based on all information available at time $t-1$. These forecasts are defined recursively by \citep[see][]{Silver2017,Axsater2015}:
\begin{align}
\label{eq:ses}
    f_{t+1}& = \alpha \cdot d_t + (1 - \alpha) \cdot f_{t}\\\label{eq:ses-mse}
    MSE_{t+1}& = \beta \cdot (d_t - f_{t})^2 + (1 - \beta) \cdot MSE_{t}
\end{align}
Initial forecasts $f_{1}$ and $MSE_{1}$ can be obtained using any of the methods described in literature \citep{Silver2017,Axsater2015}. Note that in some versions of SES, only point forecasts are obtained via \eqref{eq:ses}, but this is insufficient for applications in supply chain management, and hence we focus on SES as given by (\ref{eq:ses}-\ref{eq:ses-mse}).   

A DGP is a stochastic (demand) process $\{D_t\}_{t\in \mathbb{N}}$ from which realizations $d_1,d_2,\ldots,$ can be efficiently computed, e.g. for purposes of applying reinforcement learning, multi-stage stochastic programming, or bootstrapping. Suppose we would like to model a DGP for obtaining future demand scenarios for some product, and suppose that SES is an appropriate forecasting method for that product. This could very well be the case as SES is extensively applied in industry \citep{Goltsos2022} and recommended in seminal inventory textbooks. Then it is reasonable to suppose that SES could be successfully applied to the generated demand realizations, or, in other words, that the DGP is \emph{consistent} with SES, in a sense to be defined next:
\begin{definition} \label{def:consistency}
Consider any DGP $\{D_t\}_{t \in \mathcal{N}}$, and consider the SES forecasting approach with given $\alpha$, $\beta$, and initial forecasts $f_{1}$, $MSE_{1}$. The DGP is consistent with the SES approach if (\ref{eq:ses}-\ref{eq:ses-mse}) applied to the first $t$ data points $d_1,...,d_t$ yields unbiased estimates $f_{t+1}$ and $MSE_{t+1}$ of the mean and variance of the (conditional) demand distribution $D_{t+1}$ in period $t+1$. That is, for any $t\in \mathbb{N}$:
\begin{align}\label{eq:consistent-mean}
\mu_{t+1}&:=\mathbb{E}[D_{t+1}|D_t=d_{t}, \ldots, D_1=d_1] = f_{t+1} \\\label{eq:consistent-var}
\sigma^2_{t+1}&:=\mathbb{E}[(D_{t+1}-\mu_{t+1})^2|D_t=d_{t}, \ldots, D_1=d_1] = MSE_{t+1}
\end{align}
\end{definition}
With this definition in hand, we are ready to discuss our DGP. 

\subsection{Demand generating process}\label{subsec:demand-updating}
Consider any stochastic process $\{D_t\}_{t \in T}$ that is consistent with SES. Moreover, suppose that (given information on demand until $t$), $D_{t+1}$ is normally distributed. Since $\mathbb{E}[D_1]=\mu_1=f_1$ and $\mathbb{E}[(D_1-\mu_1)^2]=\sigma_1^2=MSE_1$, the distribution of $D_1$ is fully specified. We may thus generate $d_1$, which in turn enables us to compute $f_2$ and $MSE_2$ using (\ref{eq:ses}-\ref{eq:ses-mse}). Since $\mathbb{E}[D_2]=\mu_2=f_2$ and $\mathbb{E}[(D_2-\mu_2)^2]=\sigma_2^2=MSE_2$, this in turn enables us to determine the distribution of $D_2$, which enables generation of $d_2$, etc. 

Hence, under the assumption of normally distributed demands, it is relatively straightforward to obtain a DGP that is consistent with SES. However, since normal distributions are continuous and may take on negative values, this is typically not satisfactory for purposes of supply chain optimization, especially for low volume demand (see Sections~\ref{sec:introduction} and \ref{sec:literature}). Instead, we would prefer to ensure that $D_t$ takes on values in the non-negative integers. For this purpose, we propose to adopt the fitting procedure suggested by \citet{Adan1995}. Depending on the mean $\mu$ and variance $\sigma^2$, this procedure fits one in four classes of distributions defined on the non-negative integers. The distribution parameter $a$ ($a = \sigma^2 / \mu^2 - 1 / \mu$) determines which distribution class is to be fitted. In order of increasing variability, these classes are mixtures of binomials, Poisson, mixtures of negative-binomial and mixtures of geometric distributions with balanced means. Using this fitting procedure to obtain the distribution of $D_t$, instead of assuming normal distributions, yields the DGP that will be the subject of study for this paper. 

\begin{figure}[h]
\centering
\includegraphics[width=\textwidth]{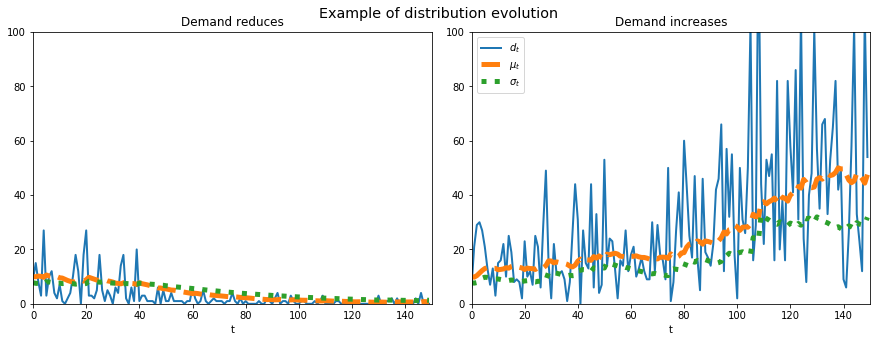}
\caption{Example of two demand sequences having identical initial parameters ($f_1 = 10, MSE_1 = 60, \alpha = \beta = 0.05$), and showing the evolution of $\mu_t, \sigma_t$ over time.}
\label{fig:example-increase-decrease}
\end{figure}

\begin{figure}[h]
\centering
\includegraphics[width=0.7\textwidth]{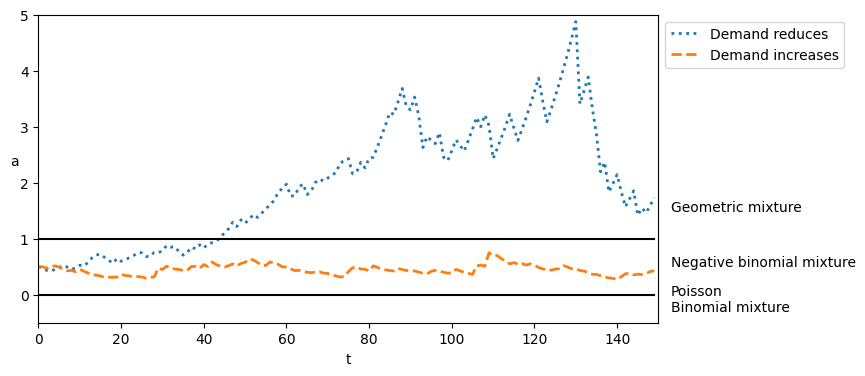}
\caption{The evolution of distribution parameter $a = \frac{\sigma_t^2}{\mu_t^2} - \frac{1}{\mu_t}$ for the two demand sequences of Figure \ref{fig:example-increase-decrease}, showing the changing distribution type over time. $f_1 = 10, MSE_1 = 60, \alpha = \beta = 0.05$}
\label{fig:a-evolution}
\end{figure}

This DGP may be used to generate sequences of discrete demands, based on an an initial estimate $f_1,MSE_1$. The resulting demand sequences are autoregressive. As an illustration, Figure {\ref{fig:example-increase-decrease}} shows two demand sequences obtained using the DGP, where in one demand sequence, the mean demand decreases, while in the other sequence the mean increases. Relatedly, if the initially provided estimate $\mu_1=f_1$ turns out to deviate considerably from $d_1$, then the uncertainty $MSE_2$ associated with $\mu_2$ is also revised upwards. 
A related and interesting feature of our proposed DGP is that the type of distribution may change when the forecast is updated over time, as the distribution parameter $a$ also changes over time as a result. This is shown in Figure {\ref{fig:a-evolution}} for the two example demand sequences of Figure {\ref{fig:example-increase-decrease}}. Note that for the demand sequence for which the mean decreases, the demand becomes intermittent, and the type of distribution that is fitted is more volatile  (i.e., a geometric mixture distribution) than the original distribution that was based on initial estimates (i.e., a negative binomial mixture). In the case of increasing demand, the distribution class remains the same. 

There is one caveat, however: it is well-known \citep[see][]{Adan1995} that there does not exist a distribution on the non-negative integers for any combination of $\mu$ and $\sigma^2$; some combinations are not \emph{feasible}. A relevant open question is then whether it is possible to define a consistent DGP on the non-negative integers. 

To see why this question is relevant, suppose that $f_1=\mu_1$ and $MSE_1=\sigma^2_1$ represent a feasible combination, such that a discrete distribution can be fitted on $D_1$. This enables the generation of $d_1$, which yields $f_2=\mu_2$ and $MSE_2=\sigma^2_2$. But under which circumstances can we \emph{guarantee} that $\mu_2,\sigma^2_2$ is again feasible, etc., such that the sequence can be completed? The next section is devoted to answering this question. 

\subsection{Consistent DGPs on the non-negative integers}\label{sec:feasibility}
As discussed in the previous section, challenges may arise when attempting to define a consistent DGP on the non-negative integers. For instance, it is impossible to specify an integral distribution with mean 0.5 and variance less than 0.25. To ensure consistency with SES, our updates to the mean and variance of $D_t$ must follow \eqref{eq:ses} and \eqref{eq:ses-mse}, which may collide with the need to arrive at feasible combinations of $\mu_t$ and $\sigma_t^2$. 

Theorem \ref{theorem:feasible} demonstrates that such collisions can be avoided when $\alpha=\beta$. In Theorem \ref{theorem:infeasible} we show that when $\alpha \neq \beta$, feasibility cannot be guaranteed. Before presenting these theorems, we first present the next Lemma, which rephrases Lemma 2.1 of \citet{Adan1995}:
\begin{lemma} \label{lemma:simple}
For a pair of non-negative, real numbers ($\mu, \sigma^2$), there exists a corresponding random variable $X$ on the non-negative integers if and only if
\begin{equation} \label{eq:lemma-equations}
 \sigma^2 \geq (\mu-\lfloor \mu \rfloor)(1-\mu+\lfloor \mu \rfloor)
\end{equation}
implying that the variance of a distribution should be at least the variance of a binomial(1,$\mu-\lfloor \mu \rfloor$) distribution.
\end{lemma}

\begin{proof} \label{proof:simple}
Following Lemma 2.1 in \cite{Adan1995}, let $k = \lfloor \mu \rfloor$. Let $\delta$ denote the difference between $k$ and $\mu$: $\delta = \mu - k = \mu - \lfloor \mu \rfloor$, meaning that $0 \leq \delta < 1$. We can rewrite $k = \mu - \delta$, and then we substitute $k$ in their Lemma 2.1:
\begin{equation}\label{eq:rewritten-lemma}
\begin{split}
\frac{\sigma^2}{\mu^2} & \geq \frac{2k + 1}{\mu} - \frac{k(k+1)}{\mu^2} - 1 \\
& \geq \frac{2(\mu - \delta) + 1}{\mu} - \frac{(\mu - \delta)(\mu-\delta+1)}{\mu^2} - 1 \\
& \geq \frac{2\mu - 2\delta + 1}{\mu} - \frac{\mu^2 - 2\delta \mu + \mu - \delta + \delta^2}{\mu^2} - 1 \\
& \geq \left(\frac{2\mu}{\mu} - \frac{2\delta}{\mu} + \frac{1}{\mu}\right) - \left(\frac{\mu^2}{\mu^2} - \frac{2\mu\delta}{\mu^2} + \frac{\mu}{\mu^2} + \frac{\delta^2}{\mu^2} - \frac{\delta}{\mu^2} \right) - 1 \\
& \geq 2 - \frac{2\delta}{\mu} + \frac{1}{\mu} - 1 + \frac{2\delta}{\mu} - \frac{1}{\mu} - \frac{\delta^2}{\mu^2} + \frac{\delta}{\mu^2} - 1 \\
& \geq -\frac{\delta^2}{\mu^2} + \frac{\delta}{\mu^2} \\
& \geq \frac{\delta(1 - \delta)}{\mu^2} \\
\sigma^2 & \geq \delta(1 - \delta)= (\mu-\lfloor \mu \rfloor)(1-\mu+\lfloor \mu \rfloor) \\
\end{split}
\end{equation}
\end{proof}

We are now ready to prove the main result of this section:
\begin{theorem} \label{theorem:feasible}
Let $\{D_t\}_{t\in \mathbb{N}}$ be a stochastic demand process on the non-negative integer domain, and suppose that $(\mu_1,\sigma_1^2)$ satisfies feasibility condition \eqref{eq:lemma-equations}. Given $d_1,\ldots,d_{t}$, let $\mu_t=f_t$, $\sigma_t^2=MSE_t$  be defined as in (\ref{eq:ses}-\ref{eq:ses-mse}), using smoothing parameters $\alpha$ and $\beta = \alpha$. Then the resulting combination ($\mu_t$, $\sigma^2_t$) satisfies feasibility condition \eqref{eq:lemma-equations}, and may hence be used to fit a new non-negative integer distribution.
\end{theorem}

\input{Proof.tex}

Whereas the stochastic demand process remains feasible to fit a non-negative integer distribution when using $\alpha = \beta$, this feasibility cannot be guaranteed if $\alpha \neq \beta$, as shown in Theorem \ref{theorem:infeasible}.

\begin{theorem} \label{theorem:infeasible}
Let $\{D_t\}_{t\in \mathbb{N}}$ be a stochastic demand process on the non-negative integer domain, and suppose that $(\mu_1,\sigma_1^2)$ satisfies feasibility condition \eqref{eq:lemma-equations}. Given $d_1,\ldots,d_{t}$, let $\mu_t=f_t$, $\sigma_t^2=MSE_t$  be defined as in (\ref{eq:ses}-\ref{eq:ses-mse}), using smoothing parameters $\alpha$ and $\beta \neq \alpha$. Then the resulting combination ($\mu_t$, $\sigma^2_t$) is not guaranteed to satisfy feasibility condition \eqref{eq:lemma-equations}, and with positive probability it cannot be used to fit a new non-negative integer distribution.
\end{theorem}

\begin{proof}
We prove that we cannot guarantee feasibility of the resulting combination of $\mu_{t+1}$ and $\sigma_{t+1}$ for $\alpha \neq \beta$ by giving a counterexample that shows that feasibility condition \eqref{eq:lemma-equations} no longer holds. In the counterexample we have a $Bin(1,p)$ distribution at time $t$, with $\mu_t = p$ and $\sigma_t^2=p(1-p)$. Suppose we use smoothing constants $\alpha$ and $\beta$ and that we draw $d_t=0$. The updated values are now \eqref{eq:counterexample-mu-updated} and \eqref{eq:counterexample-sigma-updated}.
\begin{equation}\label{eq:counterexample-mu-updated}
\mu_{t+1} = (1-\alpha)\mu_t + \alpha d_t = (1 - \alpha)p
\end{equation}
\begin{equation}\label{eq:counterexample-sigma-updated}
\sigma_{t+1}^2=(1-\beta)\sigma_t^2 + \beta (d_t - \mu_t)^2 = (1-\beta) p(1-p) +\beta p^2
\end{equation}
Note that in this binomial distribution, $\delta_{t+1} = \mu_{t+1}$. To fit a feasible non-negative integer distribution, feasibility condition \eqref{eq:lemma-equations} must hold. We find using \eqref{eq:counterexample-rewritten} that it could also be that $\sigma_{t+1}^2 - \mu_{t+1} (1-\mu_{t+1}) < 0$ when using $\alpha \neq \beta$, as $(p-2p^2)$ can take positive values as well as negative values.
\begin{equation} \label{eq:counterexample-rewritten}
\begin{split}
\sigma_{t+1}^2 - \mu_{t+1} (1-\mu_{t+1}) & = (1-\beta) p(1-p) +\beta p^2 - (1 - \alpha)p(1 - (1 - \alpha)p)\\
& = p - p \beta - p^2 + 2 \beta p^2 - (p - p^2 +2 \alpha p^2 - \alpha p - \alpha^2 p^2) \\
& = \alpha^2 p^2 + 2 \beta p^2 - 2 \alpha p^2 + \alpha p - \beta p \\
& =(\alpha p)^2+(\alpha-\beta)(p-2p^2)\\  
\end{split}
\end{equation}
In that case feasibility condition \eqref{eq:lemma-equations} does not hold, and there is no feasible non-negative integer distribution for that combination of $\mu_{t+1}$ and $\sigma_{t+1}$, thus completing our counterexample.
\end{proof}

We have now shown the necessary condition for which we can define a consistent DGP on the non-negative integers: one needs to use identical smoothing parameters for the mean and variance (\ref{eq:ses}-\ref{eq:ses-mse}). 

\subsection{Longer term demand estimates}\label{subsec:longtermestimates}
In this section, we make a few observations regarding the longer-term one-step-ahead demand expectations in the proposed DGP. As illustrated in Figure~\ref{fig:example-increase-decrease}, the DGP may yield individual demand realizations that, over time, deviate substantially from $f_1=\mu_1$ in either direction. One may wonder whether over the longer term, demand realizations from the DGP are expected to be higher or lower compared to the initial estimate. When utilizing the different demand samples in a method such as DRL, it is desirable if on average they remain in line with the initial forecast, i.e. they are unbiased. In fact, since $\mu_t$ is known, based on $d_1,\ldots,d_{t-1}$, we may observe the following:
\begin{equation}\label{eq:mean-expectation}
\begin{split}
\mathbb{E}[D_{t+1}]= \mathbb{E}[\mathbb{E}[f_{t+1}|f_t]] & =\mathbb{E}[(1 - \alpha) \cdot \mathop{\mathbb{E}}[f_t|f_t] + \alpha \cdot \mathop{\mathbb{E}}[{d_t}|f_t]] \\
& =\mathbb{E}[ (1 - \alpha) \cdot f_t + \alpha \cdot f_t] \\
& =\mathbb{E}[f_t] =\mathbb{E}[D_t] = f_1
\end{split}
\end{equation}
Here, the first equality follows from \eqref{eq:ses}, the second equality holds because $f_t=\mu_t$, and the last equality holds by continuing this type of argument recursively. Thus, over time, the expectation of the mean of the DGP remains the same as the initial estimate of the mean $f_1$. A similar identity can be derived for the expected variance:
\begin{equation}\label{eq:var-expectation-rewritten}
\begin{split}
\mathbb{E}[MSE_{t+1}] = \mathbb{E}[\mathbb{E}[MSE_{t+1}|MSE_t]] & = \mathbb{E}[(1-\beta)\cdot \mathbb{E}[MSE_t] + \beta \cdot \mathbb{E}[(d_t - f_t)^2]]\\
& = \mathbb{E}[(1-\beta) \cdot \mathbb{E}[MSE_t] + \beta \cdot \mathbb{E}[MSE_t]]\\
& = \mathbb{E}[MSE_t] = \mathbb{E}[\sigma_t^2] = MSE_1
\end{split}
\end{equation}
Here, we observe that the parameter $MSE_t$ that represents the estimated one-step-ahead variance remains in expectation equal to the initial estimate of the one-step-ahead variance $MSE_1$. 

\begin{figure}
    \centering
    \includegraphics[width = 0.7\textwidth]{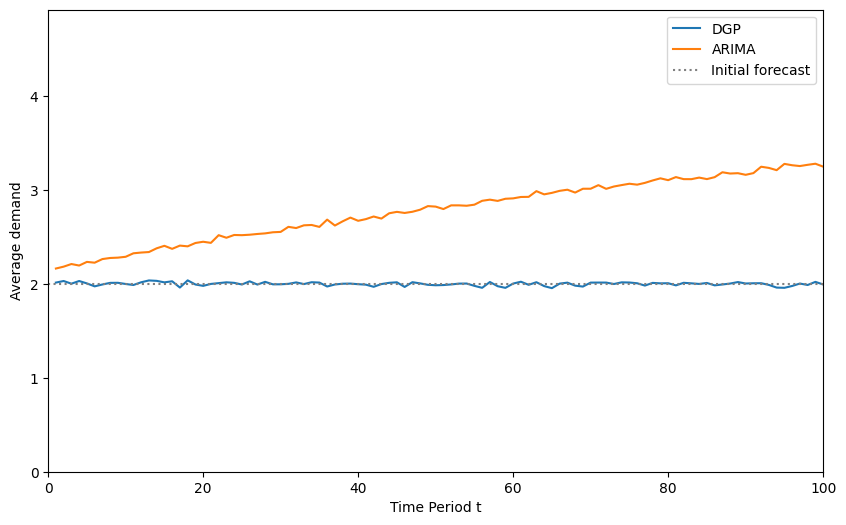}
    \caption{The average value of 10,000 demand samples at increasing time periods $t$. In this case the initial forecast is equal to 2, the standard deviation $\sigma = 2$, smoothing parameter $\alpha = 0.1$. Using the ARIMA process to generate demand introduces bias, while the DGP proposed in this paper remains unbiased.}
    \label{fig:bias-arima}
\end{figure}

In fact, using an ARIMA process for generating non-negative integer demand trajectories does not exhibit this lack of bias. If one were to use the (0,1,1) ARIMA process to generate demand observations, truncate negative values, and round the continuous values to the nearest integer, the average demand in the long-term would deviate significantly from the initial forecast, as shown in Figure \ref{fig:bias-arima}. While the ARIMA process would generally yield unbiased trajectories in cases with a very low coefficient of variation, once the standard deviation is in a similar magnitude of the mean, a lot of bias is introduced. This highlights the importance of having an appropriate DGP for non-negative integer distributions.

%% file: input/Proof.tex
\begin{proof} \label{proof:feasible}
We will prove Theorem \ref{theorem:feasible} by induction: we will show that feasibility condition \eqref{eq:lemma-equations} holds for all potential cases of $\mu_{t+1}$ and $\sigma_{t+1}^2$, given that it holds for $\mu_t$ and $\sigma_{t}^2$. In all cases, let $k_t=\lfloor \mu_t \rfloor$, and $\delta_t = \mu_t - k_t$. We distinguish the case where the demand $d_t$ is \textit{below average} $\mu_t$ (case 1) and the case where the demand $d_t$ is \textit{above average} $\mu_t$ (case 2). For case 1 and 2, we distinguish three potential sub-cases for $k_{t+1}$ and show that in each of the sub-cases, the feasibility condition \eqref{eq:lemma-equations} continues to hold, thereby proving Theorem \ref{theorem:feasible}. We show this by deriving expressions for $\sigma_{t+1}^2$ and $\delta_{t+1}$, both in terms of $\alpha$, $n$ and $\delta_t$, and then comparing them to show that $\sigma^2 \geq \delta(1-\delta)$. One observation we make based on Lemma \ref{lemma:simple} is that the maximum value of $\delta(1-\delta)$ is 0.25, so that, if the updated variance reaches that value, feasibility is guaranteed. 

\textbf{Case 1} Suppose that demand is \textit{below average}, and let $d_t=k_t-n$. Then $n \in \{0,1,\cdots,k_t\}$. When demand is below average, $k_t-n\leq k_{t+1} \leq k_t$, and then the updated parameters are given by \eqref{eq:update-mu-below-average} and \eqref{eq:update-sigma-below-average}. 
\begin{equation}\label{eq:update-mu-below-average}
\begin{split}
\mu_{t+1} & = (1 - \alpha)\mu_t + \alpha d_t = \mu_t - \alpha \mu_t + \alpha k_t - \alpha n \\
& = \mu_t - \alpha \mu_t + \alpha (\mu_t - \delta_t) - \alpha n = \mu_t - \alpha (n + \delta_t)
\end{split}
\end{equation}
\begin{equation}\label{eq:update-sigma-below-average}
\begin{split}
\sigma_{t+1}^2&=(1-\alpha)\sigma_t^2 +\alpha (d_t - \mu_t)^2 = (1-\alpha)\sigma_t^2 +\alpha (k_t - n - \mu_t)^2 \\
& = (1-\alpha)\sigma_t^2 +\alpha (\mu_t - \delta_t - n - \mu_t)^2 = (1-\alpha)\sigma_t^2 +\alpha (n + \delta_t)^2
\end{split}
\end{equation}

\underline{Case 1a}: $k_{t+1} \leq k_t-2$ \\
For $k_{t+1} \leq k_t - 2$, we need to have $\mu_t-\mu_{t+1}>1$ and $n\geq 2$. This leads to $\alpha(n+\delta_t)>1$ following \eqref{eq:update-mu-below-average}, meaning that $\alpha(n+\delta_t)^2>1$ following \eqref{eq:update-sigma-below-average}. Thus $\sigma_{t+1}^2>1$, and feasibility condition \eqref{eq:lemma-equations} holds.\\

\underline{Case 1b}: $k_{t+1} = k_t$\\
We need to derive the expressions for $\sigma_{t+1}^2$ and $\delta_{t+1}(1-\delta_{t+1})$, and then compare them to show that feasibility condition \eqref{eq:lemma-equations} holds. In \eqref{eq:case-1b-sigma} we derive the expression for $\sigma_{t+1}^2$ in terms of $\alpha$, $n$, and $\delta_t$, where in the inequality we have used the fact that $\sigma_t^2 \geq \delta_t(1 - \delta_t)$. 
\begin{equation}\label{eq:case-1b-sigma}
\begin{split}
 \sigma_{t+1}^2 & = (1 - \alpha)\sigma_t^2 + \alpha (\delta_t + n)^2 \\
 & \geq  (1 - \alpha)\delta_t(1 - \delta_t) + \alpha (\delta_t^2 + 2 \delta_t n +n^2) \\
 & = (1 - \alpha)(\delta_t - \delta_t^2) + \alpha \delta_t^2 + 2\alpha \delta_t n + \alpha n^2 \\
 & =  \delta_t - \delta_t^2 - \alpha \delta_t + 2 \alpha \delta_t^2 + 2\alpha \delta_t n + \alpha n^2 
\end{split}
\end{equation}
We derive an expression for $\delta_{t+1}$ in terms of $\alpha$, $n$ and $\delta_t$ in \eqref{eq:case-1b-delta}. 
\begin{equation}\label{eq:case-1b-delta}
\begin{split}
\delta_{t+1} &= \mu_{t+1} - k_{t+1} = \mu_t - \alpha(n + \delta_t) - k_t \\
& = \mu_t - \alpha n - \alpha \delta_t - \mu_t + \delta_t = (1 - \alpha)\delta_t - \alpha n 
\end{split}
\end{equation}
In \eqref{eq:case-1b-delta-eq} we derive the expression for $\delta_{t+1}(1-\delta_{t+1})$ in terms of $\alpha$, $n$ and $\delta_t$. 
\begin{equation}\label{eq:case-1b-delta-eq}
\begin{split}
\delta_{t+1}(1 - \delta_{t+1})  & = ((1 - \alpha)\delta_t - \alpha n)(1 - (1 - \alpha)\delta_t + \alpha n)) \\
& = (1-\alpha)\delta_t - (1-\alpha)^2\delta_t^2 + 2 \alpha(1-\alpha)\delta_t n - \alpha n - \alpha ^2 n^2 \\
& = \delta_t - \alpha \delta_t - (\delta_t^2 - 2 \alpha \delta_t^2 + \alpha^2 \delta_t^2) + 2 \alpha \delta_t n - 2\alpha^2\delta_t n - \alpha n - \alpha ^2 n^2 \\
& = \delta_t - \alpha \delta_t - \delta_t^2 + 2 \alpha \delta_t^2 - \alpha^2 \delta_t^2 + 2 \alpha \delta_t n - 2\alpha^2\delta_t n - \alpha n - \alpha ^2 n^2
\end{split}
\end{equation}
Now that we have the desired expressions of $\sigma_{t+1}^2$ \eqref{eq:case-1b-sigma} and $\delta_{t+1}(1-\delta_{t+1})$ \eqref{eq:case-1b-delta-eq} in terms of $\alpha$, $n$ and $\delta_t$, we show in \eqref{eq:case-1b} that $\sigma_{t+1}^2 - \delta_{t+1}(1-\delta_{t+1}) \geq 0$. This proves that in the case where  $k_{t+1} = k_t$, feasibility condition \eqref{eq:lemma-equations} holds.
\begin{equation}\label{eq:case-1b}
\begin{split}
\sigma_{t+1}^2 - \delta_{t+1}(1-\delta_{t+1}) & \geq \delta_t - \delta_t^2 - \alpha \delta_t + 2 \alpha \delta_t^2 + 2\alpha \delta_t n + \alpha n^2   \\
& - \delta_t + \alpha \delta_t + \delta_t^2 - 2 \alpha \delta_t^2 + \alpha^2 \delta_t^2 - 2 \alpha \delta_t n + 2\alpha^2\delta_t n + \alpha n + \alpha ^2 n^2  \\
& = \alpha n^2 + \alpha^2 \delta_t^2 + 2\alpha^2\delta_t n + \alpha n + \alpha ^2 n^2  \\
& \geq 0
\end{split}
\end{equation}

\underline{Case 1c}: $k_{t+1} = k_t - 1$\\
We again need to derive the expressions for $\sigma_{t+1}^2$ and $\delta_{t+1}(1-\delta_{t+1})$, and then compare them to show that feasibility condition \eqref{eq:lemma-equations} holds. In \eqref{eq:case-1b-sigma} we have already derived the expression for $\sigma_{t+1}^2$ in terms of $\alpha$, $n$, and $\delta_t$. We derive an expression for $\delta_{t+1}$ in terms of $\alpha$, $n$ and $\delta_t$ in \eqref{eq:case-1c-delta}. 
\begin{equation}\label{eq:case-1c-delta}
\begin{split}
\delta_{t+1} &= \mu_{t+1} - k_{t+1} = \mu_t - \alpha(n + \delta_t) - (k_t - 1) \\
& = \mu_t - \alpha n + \alpha \delta_t - \mu_t + \delta_t + 1 = (1 - \alpha)\delta_t - \alpha n + 1 \\
\end{split}
\end{equation}
In \eqref{eq:case-1c-delta-eq} we derive the expression for $\delta_{t+1}(1-\delta_{t+1})$ in terms of $\alpha$, $n$ and $\delta_t$. 
\begin{equation}\label{eq:case-1c-delta-eq}
\begin{split}
\delta_{t+1}(1 - \delta_{t+1}) & =  ((1-\alpha)\delta_t - \alpha n + 1)(1 - ((1-\alpha)\delta_t - \alpha n + 1))  \\
& =  ((1-\alpha)\delta_t - \alpha n + 1)(\alpha n - (1-\alpha)\delta_t)  \\
& =  2 \alpha(1-\alpha)\delta_t n - (1-\alpha)^2\delta_t^2 - \alpha^2 n^2 + \alpha n - (1-\alpha)\delta_t \\
& =  2 \alpha \delta_t n - 2 \alpha^2 \delta_t n - \delta_t^2 + 2 \alpha \delta_t^2 - \alpha^2 \delta_t^2 - \alpha^2 n^2 + \alpha n - \delta_t + \alpha \delta_t \\
\end{split}
\end{equation}
Now that we have the desired expressions of $\sigma_{t+1}^2$ \eqref{eq:case-1b-sigma} and $\delta_{t+1}(1-\delta_{t+1})$ \eqref{eq:case-1c-delta-eq} in terms of $\alpha$, $n$ and $\delta_t$, we show in \eqref{eq:case-1c} that $\sigma_{t+1}^2 - \delta_{t+1}(1-\delta_{t+1}) \geq 0$. This proves that in the case where  $k_{t+1} = k_t-1$, feasibility condition \eqref{eq:lemma-equations} holds.
\begin{equation}\label{eq:case-1c}
\begin{split}
\sigma_{t+1}^2 - \delta_{t+1}(1-\delta_{t+1}) & \geq \delta_t - \delta_t^2 - \alpha \delta_t + 2 \alpha \delta_t^2 + 2\alpha \delta_t n + \alpha n^2   \\
& - 2 \alpha \delta_t n + 2 \alpha^2 \delta_t n + \delta_t^2 - 2 \alpha \delta_t^2 + \alpha^2 \delta_t^2 + \alpha^2 n^2 - \alpha n + \delta_t - \alpha \delta_t  \\
& = 2\delta_t - 2 \alpha \delta_t  + \alpha n^2 + 2 \alpha^2 \delta_t n + \alpha^2 \delta_t^2 + \alpha^2 n^2 - \alpha n \\
& = 2(1 - \alpha)\delta_t  + \alpha n(n - 1)+ 2 \alpha^2 \delta_t n + \alpha^2 \delta_t^2 + \alpha^2 n^2\\
& \geq 0\\
\end{split}
\end{equation}

\textbf{Case 2} Now suppose that demand is \textit{above average}, and let $d_t=k+n$. Then $n \in \{1, 2, \cdots\}$. When demand is above average, $k_t \leq k_{t+1} \leq k_t + n$, and then the updated parameters are given by \eqref{eq:update-above-average-mu} and \eqref{eq:update-above-average-sigma}.  
\begin{equation}\label{eq:update-above-average-mu}
\begin{split}
\mu_{t+1} & = (1 - \alpha)\mu_t + \alpha d_t = \mu_t - \alpha \mu_t + \alpha k_t + \alpha n \\
& = \mu_t - \alpha \mu_t + \alpha (\mu_t - \delta_t) + \alpha n = \mu_t + \alpha (n - \delta_t)
\end{split}
\end{equation}
\begin{equation}\label{eq:update-above-average-sigma}
\begin{split}
\sigma_{t+1}^2&=(1-\alpha)\sigma_t^2 +\alpha (d_t - \mu_t)^2 = (1-\alpha)\sigma_t^2 +\alpha (k_t + n - \mu_t)^2 \\
& = (1-\alpha)\sigma_t^2 +\alpha (\mu_t - \delta_t + n - \mu_t)^2 = (1-\alpha)\sigma_t^2 +\alpha (n - \delta_t)^2
\end{split}
\end{equation}

\underline{Case 2a}: $k_{t+1} \geq k_t + 2$ \\
For $k_{t+1} \geq k_t + 2$, we need to have $\mu_t-\mu_{t+1}>1$ and $n \geq 2$. This leads to $\alpha(n-\delta_t)>1$ following \eqref{eq:update-above-average-mu}, meaning that $\alpha(n-\delta_t)^2>1$ following \eqref{eq:update-above-average-sigma}. Thus $\sigma_{t+1}^2>1$, and thus feasibility condition \eqref{eq:lemma-equations} holds.\\

\underline{Case 2b}: $k_{t+1} = k_t$ \\
We need to derive the expressions for $\sigma_{t+1}^2$ and $\delta_{t+1}(1-\delta_{t+1})$, and then compare them to show that feasibility condition \eqref{eq:lemma-equations} holds. In \eqref{eq:case-2b-sigma} we derive the expression for $\sigma_{t+1}^2$ in terms of $\alpha$, $n$ and $\delta_t$, where in the inequality we have used the fact that $\sigma_t^2 \geq \delta_t(1 - \delta_t)$. 
\begin{equation}\label{eq:case-2b-sigma}
\begin{split}
 \sigma_{t+1}^2 & = (1 - \alpha)\sigma_t^2 + \alpha (n - \delta_t)^2 \\
 & \geq  (1 - \alpha)\delta_t(1 - \delta_t) + \alpha (n^2 - 2 \delta_t n + \delta_t^2) \\
 & = \delta_t - \delta_t^2 -\alpha \delta_t + 2 \alpha \delta_t^2 + \alpha n^2 - 2 \alpha \delta_t n \\
\end{split}
\end{equation}
We derive an expression for $\delta_{t+1}$ in terms of $\alpha$, $n$ and $\delta_t$ in  \eqref{eq:case-2b-delta}.
 \begin{equation}\label{eq:case-2b-delta}
\begin{split}
\delta_{t+1} &= \mu_{t+1} - k_{t+1} = \mu_t + \alpha(n - \delta_t) - k_t \\
& = \mu_t + \alpha n - \alpha \delta_t - \mu_t + \delta_t = \alpha n + (1 - \alpha)\delta_t
\end{split}
 \end{equation}
 In \eqref{eq:case-2b-delta-eq} we derive the expression for $\delta_{t+1}(1-\delta_{t+1})$ in terms of $\alpha$, $n$ and $\delta_t$.
\begin{equation}\label{eq:case-2b-delta-eq}
\begin{split}
\delta_{t+1}(1 - \delta_{t+1}) & = (\alpha n + (1 - \alpha)\delta_t)(1 - (\alpha n + (1 - \alpha)\delta_t))\\
 & = \alpha n - \alpha^2 n^2 -  2 \alpha (1 - \alpha) \delta_t n + (1 - \alpha)\delta_t - (1-\alpha)^2\delta_t^2 \\
  & = \alpha n - \alpha^2 n^2 - 2 \alpha \delta_t n + 2 \alpha^2 \delta_t n + \delta_t - \alpha \delta_t - \delta_t^2 + 2 \alpha \delta_t^2 -\alpha^2\delta_t^2 
\end{split}
\end{equation}
Now that we have the desired expressions of $\sigma_{t+1}^2$ \eqref{eq:case-2b-sigma} and $\delta_{t+1}(1-\delta_{t+1})$ \eqref{eq:case-2b-delta-eq} in terms of $\alpha$, $n$ and $\delta_t$, we show in \eqref{eq:case-2b} that $\sigma_{t+1}^2 - \delta_{t+1}(1-\delta_{t+1}) \geq 0$. This proves that in the case where  $k_{t+1} = k_t$, feasibility condition \eqref{eq:lemma-equations} holds.
\begin{equation}\label{eq:case-2b}
\begin{split}
\sigma_{t+1}^2 - \delta_{t+1}(1-\delta_{t+1}) & \geq \delta_t - \delta_t^2 -\alpha \delta_t + 2 \alpha \delta_t^2 + \alpha n^2 - 2 \alpha \delta_t n   \\
& - \alpha n + \alpha^2 n^2 + 2 \alpha \delta_t n - 2 \alpha^2 \delta_t n - \delta_t + \alpha \delta_t + \delta_t^2 - 2 \alpha \delta_t^2 +\alpha^2\delta_t^2 \\
& = \alpha n^2 - \alpha n + \alpha^2 n^2 - 2 \alpha^2 \delta_t n + \alpha^2 \delta_t^2 \\
& = \alpha n (n - 1) + \alpha^2(n - \delta_t)^2\\
& \geq 0 \\
\end{split}
\end{equation}

\underline{Case 2c}: $k_{t+1} = k_t + 1$ \\
We again need to derive the expressions for $\sigma_{t+1}^2$ and $\delta_{t+1}(1-\delta_{t+1})$, and then compare them to show that feasibility condition \eqref{eq:lemma-equations} holds. In \eqref{eq:case-2b-sigma} we have already derived the expression for $\sigma_{t+1}^2$ in terms of $\alpha$, $n$ and $\delta_t$. We derive an expression for $\delta_{t+1}$ in terms of $\alpha$, $n$ and $\delta_t$ in \eqref{eq:case-2c-delta}. 
 \begin{equation}\label{eq:case-2c-delta}
\begin{split}
\delta_{t+1} &= \mu_{t+1} - k_{t+1} = \mu_t + \alpha(n - \delta_t) - (k_t + 1)\\
& = \mu_t + \alpha n - \alpha \delta_t - \mu_t + \delta_t - 1= \alpha n + (1 - \alpha)\delta_t - 1
\end{split}
 \end{equation}
 In \eqref{eq:case-2c-delta-eq} we derive the expression for $\delta_{t+1}(1-\delta_{t+1})$ in terms of $\alpha$, $n$ and $\delta_t$.
\begin{equation}\label{eq:case-2c-delta-eq}
\begin{split}
\delta_{t+1}(1 - \delta_{t+1}) & = (\alpha n + (1 - \alpha)\delta_t - 1)(1 - (\alpha n + (1 - \alpha)\delta_t) - 1)\\
& = (\alpha n + (1 - \alpha)\delta_t - 1)(- \alpha n - (1 - \alpha)\delta_t))\\
& = - \alpha^2 n^2 - 2\alpha (1 - \alpha) \delta_t n - (1-\alpha)^2\delta_t^2 + \alpha n + (1 - \alpha)\delta_t \\
& = - \alpha^2 n^2 - 2\alpha \delta_t n + 2\alpha^2\delta_t n - \delta_t^2 + 2\alpha \delta_t^2 - \alpha^2\delta_t^2 + \alpha n + \delta_t - \alpha\delta_t \\
\end{split}
\end{equation}
Now that we have the desired expressions of $\sigma_{t+1}^2$ \eqref{eq:case-2b-sigma} and $\delta_{t+1}(1-\delta_{t+1})$ \eqref{eq:case-2c-delta-eq} in terms of $\alpha$, $n$ and $\delta_t$, we show in \eqref{eq:case-2c} that $\sigma_{t+1}^2 - \delta_{t+1}(1-\delta_{t+1}) \geq 0$. This proves that in the case where  $k_{t+1} = k_t+1$, feasibility condition \eqref{eq:lemma-equations} holds, as in this case $n \geq 1$.
\begin{equation}\label{eq:case-2c}
\begin{split}
\sigma_{t+1}^2 - \delta_{t+1}(1-\delta_{t+1}) & \geq \delta_t - \delta_t^2 -\alpha \delta_t + 2 \alpha \delta_t^2 + \alpha n^2 - 2 \alpha \delta_t n   \\
& - \alpha^2 n^2 + 2\alpha \delta_t n - 2\alpha^2\delta_t n + \delta_t^2 - 2\alpha \delta_t^2 + \alpha^2\delta_t^2 - \alpha n - \delta_t + \alpha\delta_t  \\
& = \alpha n^2 - \alpha n + \alpha^2 n^2 - 2 \alpha^2 \delta_t n + \alpha^2 \delta_t^2 \\
& = \alpha n (n - 1) + \alpha^2(n - \delta_t)^2\\
& \geq 0 \\
\end{split}
\end{equation}
We have now proven Theorem \ref{theorem:feasible} by induction: feasibility condition \eqref{eq:lemma-equations} holds for all potential cases of $\mu_{t+1}$ and $\sigma_{t+1}^2$, given that it holds for $\mu_t$ and $\sigma_{t}^2$.
\end{proof}

%% file: input/4-ProblemDescriptionInventory.tex
To illustrate a potential application of the DGP as described in \Cref{sec:demandmodel}, we consider an inventory control problem with low volume demand. It is a common problem in industry, for example when considering spare parts inventory. In this problem we have an inventory system where we have to replenish products with non-stationary stochastic demand, such that we are able to meet customer demand immediately from stock. Replenishment orders are delivered after a deterministic lead time ($l$). At the beginning of a period, we have to place a replenishment order. This becomes available after $l$ periods. After placing an order, we observe and satisfy demand. This means that there are $l + 1$ periods of uncertain demand to cover before the replenishment order becomes available. Any demand that cannot be satisfied is backordered. The inventory transition function that describes how demand, backorders, inventory and replenishment orders relate to each other is defined by \eqref{eq:transition-function}, with a summary of the notation shown in Table \ref{table:modelnotation}. Backorders are fulfilled first, followed by any new demand in the period \eqref{eq:satisfied-demand}. The costs per period are either inventory holding costs or backorder penalty costs \eqref{eq:period-costs}.
\begin{equation}\label{eq:transition-function}
I_{t-1} - b_{t-1} + q_t = d_t + I_t - b_t
\end{equation}
\begin{equation}\label{eq:satisfied-demand}
d_t^s = \min(I_{t-1} - b_{t-1} + q_t, d_t)
\end{equation}
\begin{equation}\label{eq:period-costs}
c_t = h \cdot I_t + p \cdot b_t
\end{equation}

\input{Table-ModelNotationInventory}

We evaluate how existing solution methods for the inventory control problem perform when encountering low volume and highly variable demand, which follows our proposed DGP. Additionally we provide a new solution method that uses some information about the DGP. We describe these solution methods in \cref{subsec:solution-methods}. The setup of numerical experiments to evaluate the performance of these solution methods is specified in \cref{subsec:numerical-experiments}. Finally, the results of these numerical experiments are presented in \cref{subsec:results}.

\subsection{Solution methods}\label{subsec:solution-methods}
A typical approach to solving the inventory control problem is to follow a base-stock policy. In this policy, we need to set a base-stock level $S_t$ that covers the demand during lead-time plus some uncertainty. Every period, the replenishment order is equal to the base-stock level minus the current inventory position ($q_{t+l} = S_t - IP_t$), if the inventory position \eqref{eq:inventory-position} is below $S_t$. 
\begin{equation}\label{eq:inventory-position}
IP_t = I_{t - 1} - b_{t-1} + \sum_{t'=t}^{t+L-1}q_{t'} 
\end{equation}
Several authors have found that a state-dependent $(s, S)$-policy is optimal for this problem (see \Cref{sec:literature}) when there is Markov-modulated non-stationary demand. In our case of autoregressive non-stationary demand, such an optimal structure has not been proven, although a policy of that structure is very common to use. In any case, it still remains difficult to find the optimal base-stock levels. However, there are some well-established guidelines in determining the parameters, namely to set the base-stock with a certain service level target in mind.

There are different kinds of service level targets under consideration in inventory control problems. Often in literature, the `in-stock probability' ($P_1$-service level) is used \eqref{eq:in-stock-probability}, which is the probability of not observing a stock-out during the lead time. Another service level is the fill rate ($P_2$-service level), which is the proportion of demand that is met during lead time \eqref{eq:fill-rate}. Typically, when computing the base-stock level the `in-stock probability' target is considered, mainly for its prevalence in common textbooks and ease of use with regards to its implementation with cumulative distribution functions. 
\begin{equation}\label{eq:in-stock-probability}
\hat{P_1} = \frac{1}{T}\sum_{t \in T} P(I_{t-1} - b_{t-1} + q_t \geq d_t)
\end{equation}
\begin{equation}\label{eq:fill-rate}
\hat{P_2} = \sum_{t \in T} \frac{d_t^s}{d_t}
\end{equation}

When considering an in-stock probability target $P_1^*$, the base-stock level can be computed by $S=L \mu + \phi(P_1^*) \sigma \sqrt{L}$. A key assumption in this method is that the demand follows a (stationary) normal distribution that is i.i.d. over time, which makes the cumulative distribution function (CDF) for the standard normal random variable $\phi(P_1^*)$ applicable. Note that since we have a periodic review of the inventory (every period), the lead time that we need to consider in the base-stock level calculations is the sum of the lead time and review period ($L = l+1$). In cases that a different demand distribution is assumed, the CDF is required to determine the base-stock level. In the experiments, we use the CDF of the standard normal random variable.

The standard deviation of demand is often linked to the forecast error, and \citet{Prak2017} show that even with stationary demand, the forecast errors over the lead time demand are positively correlated. Without considering that correlation, the standard deviation of demand is often underestimated and needs to be corrected. The standard deviation may need additional corrections due to the non-stationarity of demand.

\paragraph{Solution method 1 (S1)} \citet{Graves1999} provides a correction factor in the case that demand follows an ARIMA process that is forecasted with single exponential smoothing \eqref{eq:basestock-level-graves} with parameter $\alpha$. This indicates that the standard deviation should be inflated due to the additional uncertainty in the demand process. 
\begin{equation}\label{eq:basestock-level-graves}
S_{ses}=L \mu + \phi(P_1^*) \sigma \sqrt{L} \sqrt{1 + \alpha (L - 1) + \alpha^2\frac{(L-1)(2L-1)}{6}}
\end{equation}
We consider two variants of this solution method. Commonly the estimate for $\mu$ is updated in ARIMA processes, while the estimate for $\sigma$ is considered to stay unchanged (S1a). The impact of using updated estimates of $\sigma$ in \eqref{eq:basestock-level-graves} is studied (S1b). Note that in Graves' model, the demand is real-valued and can take on negative values. This could lead to underestimating the base-stock level that is needed when demand is integer-valued and non-negative, especially in cases similar to the example in Figure \ref{fig:bias-arima}.

\paragraph{Solution method 2 (S2)} Using the DGP proposed in this paper, we can generate non-negative and integer-valued demand scenarios. With these scenarios, we can derive an empirical CDF, which we then use to find the base-stock level \eqref{eq:basestock-level-empirical}.
\begin{equation}\label{eq:basestock-level-empirical}
S_{emp}=\hat{F_n}(P_1^*)
\end{equation}
The empirical CDF is found as follows:
\begin{enumerate}
    \item Generate $n = 10,000$ demand scenarios of length $L$, given an initial estimate for $\mu$ and $\sigma$
    \item Compute lead time demand for each scenario: $\sum_{t = 0}^{L}d_t$
    \item Compute empirical CDF based on the $n$ lead time demand observations and target in-stock probability
\end{enumerate}



\subsection{Experimental setup}\label{subsec:numerical-experiments}
To evaluate the performance of the two solution methods for computing the base-stock level in low-volume spare parts environment, we consider a number of different problem parameters. For the DGP, we consider two different cases of low volume demand, that have a highly variable initial starting distributions as given by the paper of \citet{Adan1995}. We investigate three different values for the smoothing parameter. Note that 0.1 is recommended as a maximum value as by \citet{Silver2017}. Especially in cases with longer lead times, the empirical variance of lead time demand could be very different from the normal distribution or ARIMA approximations. Therefore we consider several levels of the lead time $l$. We also consider several levels of the penalty costs $p$, which are used to compute the in-stock probability target ($P_1^*=\frac{p}{p+h}$). We look at the target in-stock probabilities to identify the efficiency curve (trade off between inventory and fill rate). A summary of the experimental settings is found in Table \ref{table:experimental-settings}.

\input{Table-ExperimentalSettings.tex}

To evaluate the performance of the three solution methods, we run 10,000 simulations for 100 time periods each. The demand in the simulation is based on the DGP described in \cref{sec:demandmodel}, generated with $\alpha$. At period 1, there is no starting inventory, and we use $2 \cdot L$ periods as warm-up periods. We are interested in the performance of the different methods on the expected costs per period ($\mathbb{E}[c_t]$) and their achieved fill rate \eqref{eq:fill-rate} combined with average on hand inventory levels. Note that in setting the base-stock level, the in-stock probability target is considered. Nevertheless, the fill rate is a more common service measure in practice as it considers the magnitude of not satisfying demand as well. Small cases of lost demand are less damaging than in case of the in-stock probability, but missing large demand has a heavier impact on the fill rate. That is why we focus on the fill rate performance in the next section.

\subsection{Results}\label{subsec:results}
\input{5-Results}

%% file: input/Table-ModelNotationInventory.tex
\begin{table}[h]
\centering
\caption{Problem notation}
{\begin{tabularx}{\textwidth}{lX}
\toprule
\textbf{Parameter} & \textbf{Description} \\ 
\midrule
$l \in \mathbb{N}$ & Replenishment lead time \\
$d_t \in \mathbb{N}$ & Realized demand in period $t$ \\
$d_t^s \in \mathbb{N}$ & Demand in period $t$ that is met immediately from stock \\
$q_t \in \mathbb{N}$ & Replenishment order, created in period $t-l$, thus becoming \newline 
available in period $t$ \\
$I_t \in \mathbb{N}$ & Ending inventory on hand in period $t$ \\
$b_t \in \mathbb{N}$ & Backorders at the end of period $t$ \\
$p$ & Penalty costs per unit on backorder \\
$h$ & Holding costs per unit on inventory \\
\bottomrule
\end{tabularx}}
\label{table:modelnotation}
\end{table}

%% file: input/Table-ExperimentalSettings.tex
\begin{table}[h]
\centering
\caption{Experimental settings}
{\begin{tabular}{lll}
\toprule
\textbf{Parameter} & \textbf{Description} & \textbf{Values}\\ 
\midrule
$f_1$ & Initial mean estimate & 0.2, 0.5 \\
$\sqrt{MSE_1}$ & Initial coefficient of variation & 4.0 \\
$\alpha$ & Smoothing parameter for DGP & 0.05, 0.1, 0.15 \\
$l$ & Lead time & 1, 5, 10, 20 \\
$P_1^*$ & In stock probability target & 0.9, 0.91, $\cdots$, 0.98 \\
$h$ & Holding cost & 1 \\
\bottomrule
\end{tabular}}
\label{table:experimental-settings}
\end{table}

%% file: input/5-Results.tex
In this section we present the results of the numerical experiments. First we show a comparison of all three solution methods on costs. Then we zoom in on the efficiency curves.

\begin{figure}[h]
    \centering
    \includegraphics[width = \textwidth]{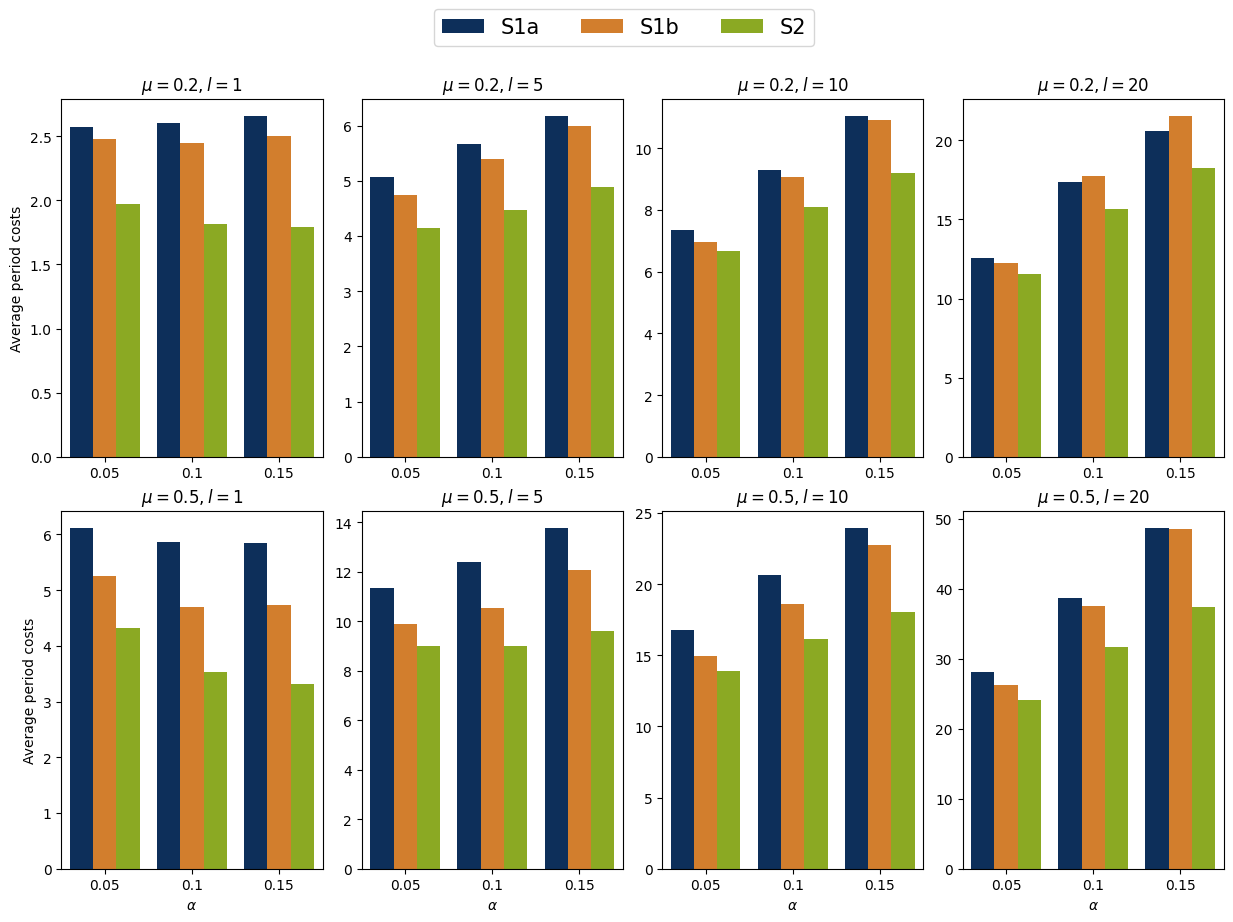}
    \caption{Average period costs for S1a, S1b, S2. $p = 9$.}
    \label{fig:period-costs}
\end{figure}

Figure \ref{fig:period-costs} shows the average period costs for S1a, S1b and S2. Our proposed method (S2) that uses the empirical distribution of the DGP outperforms existing methods on costs. Additionally, we see a large added benefit of using an updated estimate of $\sigma$ in the base stock calculations as S1b leads to lower costs than S1a. As expected, higher lead times lead to higher costs, and an increasing $\alpha$ typically has this effect as well.

\begin{figure}[h]
    \centering
    \includegraphics[width = \textwidth]{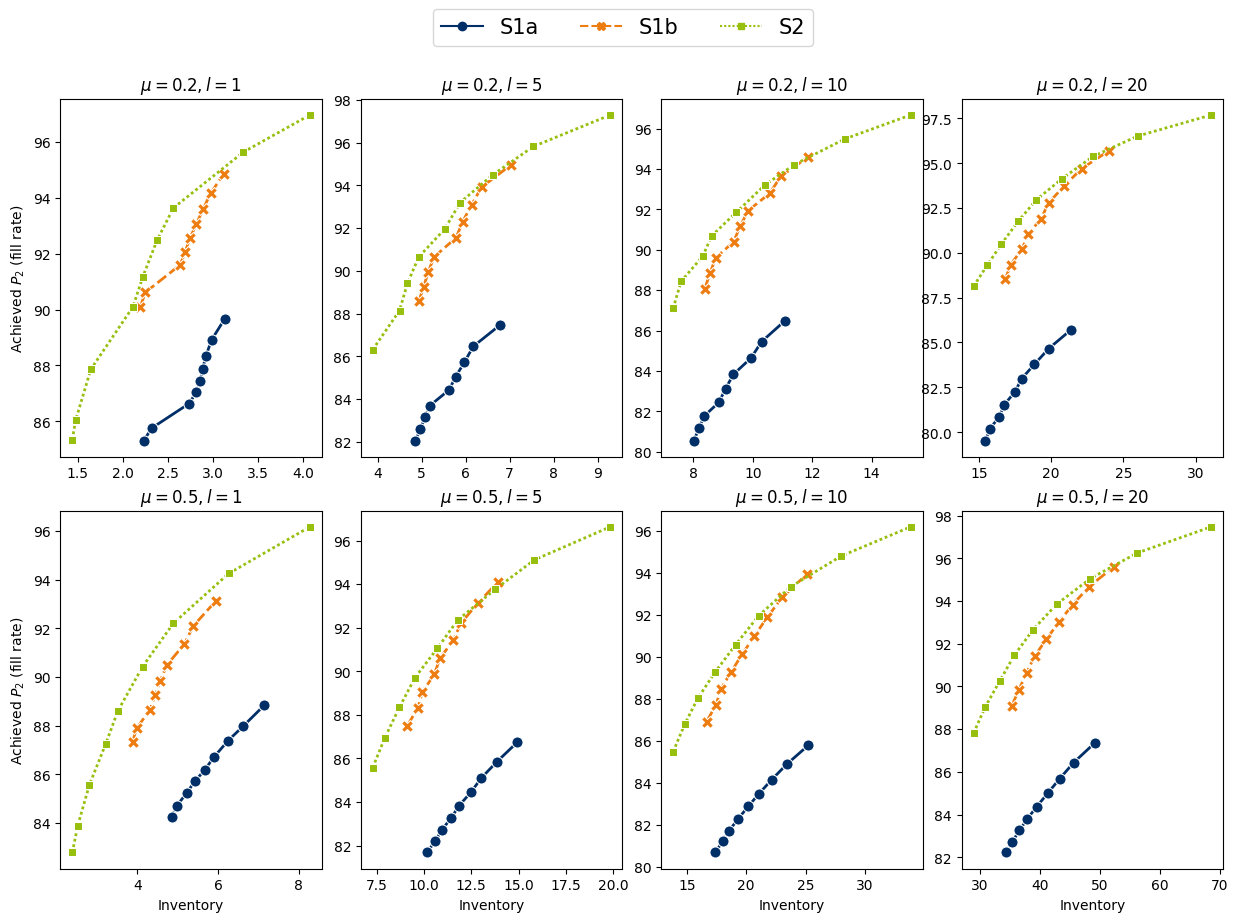}
    \caption{Efficiency curves for S1a, S1b, S2. $\alpha = 0.1$. The markers represent different service target values, where the left-most markers correspond to the lowest target.}
    \label{fig:efficiency-curve}
\end{figure}

The efficiency curve in Figure \ref{fig:efficiency-curve} demonstrates the trade off between inventory and fill rate that is achieved following each solution method. Consistent with the impact that we saw on costs, the S2 method leads to the best efficiency curve. It is always to the left of the other two methods, indicating that it can find higher service at lower levels of inventory. Especially when disregarding updates of $\sigma$ in computing the base stock level (S1a), it is nearly impossible to achieve high service levels, resulting in many backorders. Considering updated values for $\sigma$ is already a big improvement (S1b) in terms of the efficiency curve. However, using empirical distributions from the DGP (S2) leads to the best efficiency curve due to its ability to accurately grasp the non-negative and integer nature of the demand.

%% file: input/6-Conclusion.tex
We proved the existence of a demand process that is consistent with SES on the non-negative integers, and derive the required conditions. We proposed an intuitive and simple DGP based on these conditions, and demonstrated that the model is able to generate realistic and practically relevant demand scenarios. In this DGP, the expectation and variance of demand are treated as a forecast rather than an exogenously given number. This is in line with recommendations in seminal literature \citep{Axsater2015, Silver2017}. Additionally, this DGP generates non-negative integer demand, using a powerful fitting technique proposed by \citet{Adan1995}. 

Subsequently, we investigated a spare parts inventory control problem in this non-stationary demand setting using the DGP to generate the evaluation scenarios. We evaluated two existing base-stock policies, and suggested a new base-stock level computation method that utilizes the DGP to find an empirical CDF of demand during lead time, and then sets appropriate base-stock levels, outperforming existing base-stock policies. In cases these cases of low-volume demand, where the degree of non-stationarity is high, the variance is high and lead times are long, the method of \citet{Graves1999}, which considers potentially negative and continuous demand, leads to underestimations of the heavy tail of the demand distribution. The newly suggested method is better able to estimate the heavy tail of the demand distribution, an essential ingredient for finding appropriate inventory control parameters.

The characteristics and properties of the DGP in this paper can be utilized in more ways to specify and evaluate new inventory control policies for non-stationary demand settings. In addition to the inventory control problem, other supply chain operations planning problems could also be studied using the DGP in this paper. Another direction for future research is to study how trend and seasonality can also be incorporated in the DGP, as these will increase the practical relevance of the DGP even further. A final interesting future research direction is to understand how these non-stationary demand scenarios impact results found in optimization algorithms like multi-stage stochastic programming or Deep Reinforcement Learning. The non-stationarity of the scenarios could pose extra challenges to these algorithms, which requires additional research. 


%% file: main.bbl
\begin{thebibliography}{}

\bibitem [\protect \citeauthoryear {%
Adan%
, van Eenige%
\BCBL {}\ \BBA {} Resing%
}{%
Adan%
\ \protect \BOthers {.}}{%
{\protect \APACyear {1995}}%
}]{%
Adan1995}
\APACinsertmetastar {%
Adan1995}%
\begin{APACrefauthors}%
Adan, I.%
, van Eenige, M.%
\BCBL {}\ \BBA {} Resing, J.%
\end{APACrefauthors}%
\unskip\
\newblock
\APACrefYearMonthDay{1995}{}{}.
\newblock
{\BBOQ}\APACrefatitle {Fitting discrete distributions on the first two moments} {Fitting discrete distributions on the first two moments}.{\BBCQ}
\newblock
\APACjournalVolNumPages{Probability in the engineering and informational sciences}{9}{4}{623--632}.
\PrintBackRefs{\CurrentBib}

\bibitem [\protect \citeauthoryear {%
Altendorfer%
, Felberbauer%
\BCBL {}\ \BBA {} Jodlbauer%
}{%
Altendorfer%
\ \protect \BOthers {.}}{%
{\protect \APACyear {2016}}%
}]{%
Altendorfer2016}
\APACinsertmetastar {%
Altendorfer2016}%
\begin{APACrefauthors}%
Altendorfer, K.%
, Felberbauer, T.%
\BCBL {}\ \BBA {} Jodlbauer, H.%
\end{APACrefauthors}%
\unskip\
\newblock
\APACrefYearMonthDay{2016}{}{}.
\newblock
{\BBOQ}\APACrefatitle {Effects of forecast errors on optimal utilisation in aggregate production planning with stochastic customer demand} {Effects of forecast errors on optimal utilisation in aggregate production planning with stochastic customer demand}.{\BBCQ}
\newblock
\APACjournalVolNumPages{International Journal of Production Research}{54}{}{3718-3735}.
\newblock
\begin{APACrefURL} \url{http://dx.doi.org/10.1080/00207543.2016.1162918} \end{APACrefURL}
\newblock
\begin{APACrefDOI} \doi{10.1080/00207543.2016.1162918} \end{APACrefDOI}
\PrintBackRefs{\CurrentBib}

\bibitem [\protect \citeauthoryear {%
Amniattalab%
, Frenk%
\BCBL {}\ \BBA {} Hekimo{\u{g}}lu%
}{%
Amniattalab%
\ \protect \BOthers {.}}{%
{\protect \APACyear {2023}}%
}]{%
Amniattalab2023}
\APACinsertmetastar {%
Amniattalab2023}%
\begin{APACrefauthors}%
Amniattalab, A.%
, Frenk, J.%
\BCBL {}\ \BBA {} Hekimo{\u{g}}lu, M.%
\end{APACrefauthors}%
\unskip\
\newblock
\APACrefYearMonthDay{2023}{}{}.
\newblock
{\BBOQ}\APACrefatitle {On spare parts demand and the installed base concept: A theoretical approach} {On spare parts demand and the installed base concept: A theoretical approach}.{\BBCQ}
\newblock
\APACjournalVolNumPages{International Journal of Production Economics}{266}{}{}.
\PrintBackRefs{\CurrentBib}

\bibitem [\protect \citeauthoryear {%
Axs{\"{a}}ter%
}{%
Axs{\"{a}}ter%
}{%
{\protect \APACyear {2015}}%
}]{%
Axsater2015}
\APACinsertmetastar {%
Axsater2015}%
\begin{APACrefauthors}%
Axs{\"{a}}ter, S.%
\end{APACrefauthors}%
\unskip\
\newblock
\APACrefYearMonthDay{2015}{}{}.
\newblock
{\BBOQ}\APACrefatitle {{Forecasting}} {{Forecasting}}.{\BBCQ}
\newblock
\BIn{} \APACrefbtitle {Inventory Control} {Inventory control}\ (\PrintOrdinal{Third Edit}\ \BEd, \BPGS\ 7--36).
\newblock
\APACaddressPublisher{}{Springer}.
\PrintBackRefs{\CurrentBib}

\bibitem [\protect \citeauthoryear {%
{Babai}%
, Dai%
, Li%
, Syntetos%
\BCBL {}\ \BBA {} Wang%
}{%
{Babai}%
\ \protect \BOthers {.}}{%
{\protect \APACyear {2022}}%
}]{%
ZiedBabai2022}
\APACinsertmetastar {%
ZiedBabai2022}%
\begin{APACrefauthors}%
{Babai}, M\BPBI Z.%
, Dai, Y.%
, Li, Q.%
, Syntetos, A.%
\BCBL {}\ \BBA {} Wang, X.%
\end{APACrefauthors}%
\unskip\
\newblock
\APACrefYearMonthDay{2022}{}{}.
\newblock
{\BBOQ}\APACrefatitle {{Forecasting of lead-time demand variance: Implications for safety stock calculations}} {{Forecasting of lead-time demand variance: Implications for safety stock calculations}}.{\BBCQ}
\newblock
\APACjournalVolNumPages{European Journal of Operational Research}{296}{3}{846--861}.
\newblock
\begin{APACrefURL} \url{https://doi.org/10.1016/j.ejor.2021.04.017} \end{APACrefURL}
\newblock
\begin{APACrefDOI} \doi{10.1016/j.ejor.2021.04.017} \end{APACrefDOI}
\PrintBackRefs{\CurrentBib}

\bibitem [\protect \citeauthoryear {%
Bayraktar%
\ \BBA {} Ludkovski%
}{%
Bayraktar%
\ \BBA {} Ludkovski%
}{%
{\protect \APACyear {2010}}%
}]{%
Bayraktar2010}
\APACinsertmetastar {%
Bayraktar2010}%
\begin{APACrefauthors}%
Bayraktar, E.%
\BCBT {}\ \BBA {} Ludkovski, M.%
\end{APACrefauthors}%
\unskip\
\newblock
\APACrefYearMonthDay{2010}{}{}.
\newblock
{\BBOQ}\APACrefatitle {{Inventory management with partially observed nonstationary demand}} {{Inventory management with partially observed nonstationary demand}}.{\BBCQ}
\newblock
\APACjournalVolNumPages{Annals of Operations Research}{176}{1}{7--39}.
\newblock
\begin{APACrefDOI} \doi{10.1007/s10479-009-0513-8} \end{APACrefDOI}
\PrintBackRefs{\CurrentBib}

\bibitem [\protect \citeauthoryear {%
Bertsimas%
, Sim%
\BCBL {}\ \BBA {} Zhang%
}{%
Bertsimas%
\ \protect \BOthers {.}}{%
{\protect \APACyear {2019}}%
}]{%
Bertsimas2019}
\APACinsertmetastar {%
Bertsimas2019}%
\begin{APACrefauthors}%
Bertsimas, D.%
, Sim, M.%
\BCBL {}\ \BBA {} Zhang, M.%
\end{APACrefauthors}%
\unskip\
\newblock
\APACrefYearMonthDay{2019}{}{}.
\newblock
{\BBOQ}\APACrefatitle {{Adaptive distributionally robust optimization}} {{Adaptive distributionally robust optimization}}.{\BBCQ}
\newblock
\APACjournalVolNumPages{Management Science}{65}{2}{604--618}.
\newblock
\begin{APACrefDOI} \doi{10.1287/mnsc.2017.2952} \end{APACrefDOI}
\PrintBackRefs{\CurrentBib}

\bibitem [\protect \citeauthoryear {%
Boute%
, Gijsbrechts%
, Van~Jaarsveld%
\BCBL {}\ \BBA {} Vanvuchelen%
}{%
Boute%
\ \protect \BOthers {.}}{%
{\protect \APACyear {2022}}%
}]{%
boute2022deep}
\APACinsertmetastar {%
boute2022deep}%
\begin{APACrefauthors}%
Boute, R\BPBI N.%
, Gijsbrechts, J.%
, Van~Jaarsveld, W.%
\BCBL {}\ \BBA {} Vanvuchelen, N.%
\end{APACrefauthors}%
\unskip\
\newblock
\APACrefYearMonthDay{2022}{}{}.
\newblock
{\BBOQ}\APACrefatitle {Deep reinforcement learning for inventory control: A roadmap} {Deep reinforcement learning for inventory control: A roadmap}.{\BBCQ}
\newblock
\APACjournalVolNumPages{European Journal of Operational Research}{298}{2}{401--412}.
\PrintBackRefs{\CurrentBib}

\bibitem [\protect \citeauthoryear {%
Boylan%
\ \BBA {} Babai%
}{%
Boylan%
\ \BBA {} Babai%
}{%
{\protect \APACyear {2022}}%
}]{%
Boylan2022}
\APACinsertmetastar {%
Boylan2022}%
\begin{APACrefauthors}%
Boylan, J\BPBI E.%
\BCBT {}\ \BBA {} Babai, M\BPBI Z.%
\end{APACrefauthors}%
\unskip\
\newblock
\APACrefYearMonthDay{2022}{}{}.
\newblock
{\BBOQ}\APACrefatitle {{Estimating the cumulative distribution function of lead-time demand using bootstrapping with and without replacement}} {{Estimating the cumulative distribution function of lead-time demand using bootstrapping with and without replacement}}.{\BBCQ}
\newblock
\APACjournalVolNumPages{International Journal of Production Economics}{252}{April}{108586}.
\newblock
\begin{APACrefURL} \url{https://doi.org/10.1016/j.ijpe.2022.108586} \end{APACrefURL}
\newblock
\begin{APACrefDOI} \doi{10.1016/j.ijpe.2022.108586} \end{APACrefDOI}
\PrintBackRefs{\CurrentBib}

\bibitem [\protect \citeauthoryear {%
Cao%
\ \BBA {} Shen%
}{%
Cao%
\ \BBA {} Shen%
}{%
{\protect \APACyear {2019}}%
}]{%
Cao2019}
\APACinsertmetastar {%
Cao2019}%
\begin{APACrefauthors}%
Cao, Y.%
\BCBT {}\ \BBA {} Shen, Z\BPBI J\BPBI M.%
\end{APACrefauthors}%
\unskip\
\newblock
\APACrefYearMonthDay{2019}{}{}.
\newblock
{\BBOQ}\APACrefatitle {{Quantile forecasting and data-driven inventory management under nonstationary demand}} {{Quantile forecasting and data-driven inventory management under nonstationary demand}}.{\BBCQ}
\newblock
\APACjournalVolNumPages{Operations Research Letters}{47}{6}{465--472}.
\newblock
\begin{APACrefURL} \url{https://doi.org/10.1016/j.orl.2019.08.008} \end{APACrefURL}
\newblock
\begin{APACrefDOI} \doi{10.1016/j.orl.2019.08.008} \end{APACrefDOI}
\PrintBackRefs{\CurrentBib}

\bibitem [\protect \citeauthoryear {%
Gelper%
, Fried%
\BCBL {}\ \BBA {} Croux%
}{%
Gelper%
\ \protect \BOthers {.}}{%
{\protect \APACyear {2010}}%
}]{%
Gelper2010}
\APACinsertmetastar {%
Gelper2010}%
\begin{APACrefauthors}%
Gelper, S.%
, Fried, R.%
\BCBL {}\ \BBA {} Croux, C.%
\end{APACrefauthors}%
\unskip\
\newblock
\APACrefYearMonthDay{2010}{}{}.
\newblock
{\BBOQ}\APACrefatitle {{Robust forecasting with exponential and holt-winters smoothing}} {{Robust forecasting with exponential and holt-winters smoothing}}.{\BBCQ}
\newblock
\APACjournalVolNumPages{Journal of Forecasting}{29}{3}{285--300}.
\newblock
\begin{APACrefDOI} \doi{10.1002/for.1125} \end{APACrefDOI}
\PrintBackRefs{\CurrentBib}

\bibitem [\protect \citeauthoryear {%
Gilbert%
}{%
Gilbert%
}{%
{\protect \APACyear {2005}}%
}]{%
Gilbert2005}
\APACinsertmetastar {%
Gilbert2005}%
\begin{APACrefauthors}%
Gilbert, K.%
\end{APACrefauthors}%
\unskip\
\newblock
\APACrefYearMonthDay{2005}{}{}.
\newblock
{\BBOQ}\APACrefatitle {{An ARIMA supply chain model}} {{An ARIMA supply chain model}}.{\BBCQ}
\newblock
\APACjournalVolNumPages{Management Science}{51}{2}{305--310}.
\newblock
\begin{APACrefDOI} \doi{10.1287/mnsc.1040.0308} \end{APACrefDOI}
\PrintBackRefs{\CurrentBib}

\bibitem [\protect \citeauthoryear {%
Goltsos%
, Syntetos%
, Glock%
\BCBL {}\ \BBA {} Ioannou%
}{%
Goltsos%
\ \protect \BOthers {.}}{%
{\protect \APACyear {2022}}%
}]{%
Goltsos2022}
\APACinsertmetastar {%
Goltsos2022}%
\begin{APACrefauthors}%
Goltsos, T\BPBI E.%
, Syntetos, A\BPBI A.%
, Glock, C\BPBI H.%
\BCBL {}\ \BBA {} Ioannou, G.%
\end{APACrefauthors}%
\unskip\
\newblock
\APACrefYearMonthDay{2022}{}{}.
\newblock
{\BBOQ}\APACrefatitle {{Inventory – forecasting: Mind the gap}} {{Inventory – forecasting: Mind the gap}}.{\BBCQ}
\newblock
\APACjournalVolNumPages{European Journal of Operational Research}{299}{2}{397--419}.
\newblock
\begin{APACrefURL} \url{https://doi.org/10.1016/j.ejor.2021.07.040} \end{APACrefURL}
\newblock
\begin{APACrefDOI} \doi{10.1016/j.ejor.2021.07.040} \end{APACrefDOI}
\PrintBackRefs{\CurrentBib}

\bibitem [\protect \citeauthoryear {%
Graves%
}{%
Graves%
}{%
{\protect \APACyear {1999}}%
}]{%
Graves1999}
\APACinsertmetastar {%
Graves1999}%
\begin{APACrefauthors}%
Graves, S\BPBI C.%
\end{APACrefauthors}%
\unskip\
\newblock
\APACrefYearMonthDay{1999}{}{}.
\newblock
{\BBOQ}\APACrefatitle {{A Single-Item Inventory Model for a Nonstationary Demand Process}} {{A Single-Item Inventory Model for a Nonstationary Demand Process}}.{\BBCQ}
\newblock
\APACjournalVolNumPages{Manufacturing {\&} Service Operations Management}{1}{1}{50--61}.
\newblock
\begin{APACrefDOI} \doi{10.1287/msom.1.2.174} \end{APACrefDOI}
\PrintBackRefs{\CurrentBib}

\bibitem [\protect \citeauthoryear {%
Hu%
, Zhang%
\BCBL {}\ \BBA {} Zhu%
}{%
Hu%
\ \protect \BOthers {.}}{%
{\protect \APACyear {2016}}%
}]{%
Hu2016}
\APACinsertmetastar {%
Hu2016}%
\begin{APACrefauthors}%
Hu, J.%
, Zhang, C.%
\BCBL {}\ \BBA {} Zhu, C.%
\end{APACrefauthors}%
\unskip\
\newblock
\APACrefYearMonthDay{2016}{}{}.
\newblock
{\BBOQ}\APACrefatitle {{(s, S) Inventory systems with correlated demands}} {{(s, S) Inventory systems with correlated demands}}.{\BBCQ}
\newblock
\APACjournalVolNumPages{INFORMS Journal on Computing}{28}{4}{603--611}.
\PrintBackRefs{\CurrentBib}

\bibitem [\protect \citeauthoryear {%
Hyndman%
\ \BBA {} Athanasopoulos%
}{%
Hyndman%
\ \BBA {} Athanasopoulos%
}{%
{\protect \APACyear {2018}}%
}]{%
hyndman2018forecasting}
\APACinsertmetastar {%
hyndman2018forecasting}%
\begin{APACrefauthors}%
Hyndman, R\BPBI J.%
\BCBT {}\ \BBA {} Athanasopoulos, G.%
\end{APACrefauthors}%
\unskip\
\newblock
\APACrefYear{2018}.
\newblock
\APACrefbtitle {Forecasting: principles and practice} {Forecasting: principles and practice}.
\newblock
\APACaddressPublisher{}{OTexts}.
\PrintBackRefs{\CurrentBib}

\bibitem [\protect \citeauthoryear {%
Janssen%
, Strijbosch%
\BCBL {}\ \BBA {} Brekelmans%
}{%
Janssen%
\ \protect \BOthers {.}}{%
{\protect \APACyear {2009}}%
}]{%
Janssen2009}
\APACinsertmetastar {%
Janssen2009}%
\begin{APACrefauthors}%
Janssen, E.%
, Strijbosch, L.%
\BCBL {}\ \BBA {} Brekelmans, R.%
\end{APACrefauthors}%
\unskip\
\newblock
\APACrefYearMonthDay{2009}{}{}.
\newblock
{\BBOQ}\APACrefatitle {{Assessing the effects of using demand parameters estimates in inventory control and improving the performance using a correction function}} {{Assessing the effects of using demand parameters estimates in inventory control and improving the performance using a correction function}}.{\BBCQ}
\newblock
\APACjournalVolNumPages{International Journal of Production Economics}{118}{1}{34--42}.
\newblock
\begin{APACrefDOI} \doi{10.1016/j.ijpe.2008.08.029} \end{APACrefDOI}
\PrintBackRefs{\CurrentBib}

\bibitem [\protect \citeauthoryear {%
Kataoka%
}{%
Kataoka%
}{%
{\protect \APACyear {1963}}%
}]{%
kataoka1963stochastic}
\APACinsertmetastar {%
kataoka1963stochastic}%
\begin{APACrefauthors}%
Kataoka, S.%
\end{APACrefauthors}%
\unskip\
\newblock
\APACrefYearMonthDay{1963}{}{}.
\newblock
{\BBOQ}\APACrefatitle {A stochastic programming model} {A stochastic programming model}.{\BBCQ}
\newblock
\APACjournalVolNumPages{Econometrica: Journal of the Econometric Society}{}{}{181--196}.
\PrintBackRefs{\CurrentBib}

\bibitem [\protect \citeauthoryear {%
Kaya%
\ \BBA {} Ghahroodi%
}{%
Kaya%
\ \BBA {} Ghahroodi%
}{%
{\protect \APACyear {2018}}%
}]{%
Kaya2018}
\APACinsertmetastar {%
Kaya2018}%
\begin{APACrefauthors}%
Kaya, O.%
\BCBT {}\ \BBA {} Ghahroodi, S\BPBI R.%
\end{APACrefauthors}%
\unskip\
\newblock
\APACrefYearMonthDay{2018}{}{}.
\newblock
{\BBOQ}\APACrefatitle {Inventory control and pricing for perishable products under age and price dependent stochastic demand} {Inventory control and pricing for perishable products under age and price dependent stochastic demand}.{\BBCQ}
\newblock
\APACjournalVolNumPages{Mathematical Methods of Operations Research}{88}{}{}.
\PrintBackRefs{\CurrentBib}

\bibitem [\protect \citeauthoryear {%
Kolassa%
}{%
Kolassa%
}{%
{\protect \APACyear {2016}}%
}]{%
Kolassa2016}
\APACinsertmetastar {%
Kolassa2016}%
\begin{APACrefauthors}%
Kolassa, S.%
\end{APACrefauthors}%
\unskip\
\newblock
\APACrefYearMonthDay{2016}{}{}.
\newblock
{\BBOQ}\APACrefatitle {{Evaluating predictive count data distributions in retail sales forecasting}} {{Evaluating predictive count data distributions in retail sales forecasting}}.{\BBCQ}
\newblock
\APACjournalVolNumPages{International Journal of Forecasting}{32}{3}{788--803}.
\newblock
\begin{APACrefURL} \url{http://dx.doi.org/10.1016/j.ijforecast.2015.12.004} \end{APACrefURL}
\newblock
\begin{APACrefDOI} \doi{10.1016/j.ijforecast.2015.12.004} \end{APACrefDOI}
\PrintBackRefs{\CurrentBib}

\bibitem [\protect \citeauthoryear {%
Ma%
, Rossi%
\BCBL {}\ \BBA {} Archibald%
}{%
Ma%
\ \protect \BOthers {.}}{%
{\protect \APACyear {2022}}%
}]{%
Ma2022}
\APACinsertmetastar {%
Ma2022}%
\begin{APACrefauthors}%
Ma, X.%
, Rossi, R.%
\BCBL {}\ \BBA {} Archibald, T\BPBI W.%
\end{APACrefauthors}%
\unskip\
\newblock
\APACrefYearMonthDay{2022}{}{}.
\newblock
{\BBOQ}\APACrefatitle {{Approximations for non-stationary stochastic lot-sizing under (s,Q)-type policy}} {{Approximations for non-stationary stochastic lot-sizing under (s,Q)-type policy}}.{\BBCQ}
\newblock
\APACjournalVolNumPages{European Journal of Operational Research}{298}{2}{573--584}.
\newblock
\begin{APACrefURL} \url{https://doi.org/10.1016/j.ejor.2021.06.013} \end{APACrefURL}
\newblock
\begin{APACrefDOI} \doi{10.1016/j.ejor.2021.06.013} \end{APACrefDOI}
\PrintBackRefs{\CurrentBib}

\bibitem [\protect \citeauthoryear {%
Prak%
\ \BBA {} Teunter%
}{%
Prak%
\ \BBA {} Teunter%
}{%
{\protect \APACyear {2019}}%
}]{%
Prak2019}
\APACinsertmetastar {%
Prak2019}%
\begin{APACrefauthors}%
Prak, D.%
\BCBT {}\ \BBA {} Teunter, R.%
\end{APACrefauthors}%
\unskip\
\newblock
\APACrefYearMonthDay{2019}{}{}.
\newblock
{\BBOQ}\APACrefatitle {{A general method for addressing forecasting uncertainty in inventory models}} {{A general method for addressing forecasting uncertainty in inventory models}}.{\BBCQ}
\newblock
\APACjournalVolNumPages{International Journal of Forecasting}{35}{1}{224--238}.
\newblock
\begin{APACrefURL} \url{https://doi.org/10.1016/j.ijforecast.2017.11.004} \end{APACrefURL}
\newblock
\begin{APACrefDOI} \doi{10.1016/j.ijforecast.2017.11.004} \end{APACrefDOI}
\PrintBackRefs{\CurrentBib}

\bibitem [\protect \citeauthoryear {%
Prak%
, Teunter%
\BCBL {}\ \BBA {} Syntetos%
}{%
Prak%
\ \protect \BOthers {.}}{%
{\protect \APACyear {2017}}%
}]{%
Prak2017}
\APACinsertmetastar {%
Prak2017}%
\begin{APACrefauthors}%
Prak, D.%
, Teunter, R.%
\BCBL {}\ \BBA {} Syntetos, A.%
\end{APACrefauthors}%
\unskip\
\newblock
\APACrefYearMonthDay{2017}{}{}.
\newblock
{\BBOQ}\APACrefatitle {{On the calculation of safety stocks when demand is forecasted}} {{On the calculation of safety stocks when demand is forecasted}}.{\BBCQ}
\newblock
\APACjournalVolNumPages{European Journal of Operational Research}{256}{2}{454--461}.
\newblock
\begin{APACrefURL} \url{http://dx.doi.org/10.1016/j.ejor.2016.06.035} \end{APACrefURL}
\newblock
\begin{APACrefDOI} \doi{10.1016/j.ejor.2016.06.035} \end{APACrefDOI}
\PrintBackRefs{\CurrentBib}

\bibitem [\protect \citeauthoryear {%
Rostami-Tabar%
\ \BBA {} Disney%
}{%
Rostami-Tabar%
\ \BBA {} Disney%
}{%
{\protect \APACyear {2023}}%
}]{%
Rostami-Tabar2023}
\APACinsertmetastar {%
Rostami-Tabar2023}%
\begin{APACrefauthors}%
Rostami-Tabar, B.%
\BCBT {}\ \BBA {} Disney, S\BPBI M.%
\end{APACrefauthors}%
\unskip\
\newblock
\APACrefYearMonthDay{2023}{}{}.
\newblock
{\BBOQ}\APACrefatitle {{On the order-up-to policy with intermittent integer demand and logically consistent forecasts}} {{On the order-up-to policy with intermittent integer demand and logically consistent forecasts}}.{\BBCQ}
\newblock
\APACjournalVolNumPages{International Journal of Production Economics}{257}{December 2022}{108763}.
\newblock
\begin{APACrefURL} \url{https://doi.org/10.1016/j.ijpe.2022.108763} \end{APACrefURL}
\newblock
\begin{APACrefDOI} \doi{10.1016/j.ijpe.2022.108763} \end{APACrefDOI}
\PrintBackRefs{\CurrentBib}

\bibitem [\protect \citeauthoryear {%
Silver%
, Pyke%
\BCBL {}\ \BBA {} Thomas%
}{%
Silver%
\ \protect \BOthers {.}}{%
{\protect \APACyear {2017}}%
}]{%
Silver2017}
\APACinsertmetastar {%
Silver2017}%
\begin{APACrefauthors}%
Silver, E\BPBI A.%
, Pyke, D\BPBI F.%
\BCBL {}\ \BBA {} Thomas, D\BPBI J.%
\end{APACrefauthors}%
\unskip\
\newblock
\APACrefYearMonthDay{2017}{}{}.
\newblock
{\BBOQ}\APACrefatitle {{Forecasting Models and Techniques}} {{Forecasting Models and Techniques}}.{\BBCQ}
\newblock
\BIn{} \APACrefbtitle {Inventory and Production Management in Supply Chains} {Inventory and production management in supply chains}\ (\PrintOrdinal{Fourth Edition}\ \BEd, \BPGS\ 73--144).
\newblock
\APACaddressPublisher{}{Taylor {\&} Francis Group}.
\PrintBackRefs{\CurrentBib}

\bibitem [\protect \citeauthoryear {%
Snyder%
, Koehler%
\BCBL {}\ \BBA {} Ord%
}{%
Snyder%
\ \protect \BOthers {.}}{%
{\protect \APACyear {2002}}%
}]{%
Snyder2002}
\APACinsertmetastar {%
Snyder2002}%
\begin{APACrefauthors}%
Snyder, R\BPBI D.%
, Koehler, A\BPBI B.%
\BCBL {}\ \BBA {} Ord, J\BPBI K.%
\end{APACrefauthors}%
\unskip\
\newblock
\APACrefYearMonthDay{2002}{}{}.
\newblock
{\BBOQ}\APACrefatitle {{Forecasting for inventory control with exponential smoothing}} {{Forecasting for inventory control with exponential smoothing}}.{\BBCQ}
\newblock
\APACjournalVolNumPages{International Journal of Forecasting}{18}{1}{5--18}.
\newblock
\begin{APACrefDOI} \doi{10.1016/S0169-2070(01)00109-1} \end{APACrefDOI}
\PrintBackRefs{\CurrentBib}

\bibitem [\protect \citeauthoryear {%
Song%
\ \BBA {} Zipkin%
}{%
Song%
\ \BBA {} Zipkin%
}{%
{\protect \APACyear {1993}}%
}]{%
Song1993}
\APACinsertmetastar {%
Song1993}%
\begin{APACrefauthors}%
Song, J\BHBI s.%
\BCBT {}\ \BBA {} Zipkin, P.%
\end{APACrefauthors}%
\unskip\
\newblock
\APACrefYearMonthDay{1993}{}{}.
\newblock
{\BBOQ}\APACrefatitle {{Inventory Control in a Fluctuating Demand Environment}} {{Inventory Control in a Fluctuating Demand Environment}}.{\BBCQ}
\newblock
\APACjournalVolNumPages{Operations Research}{41}{2}{351--370}.
\PrintBackRefs{\CurrentBib}

\bibitem [\protect \citeauthoryear {%
Strijbosch%
, Syntetos%
, Boylan%
\BCBL {}\ \BBA {} Janssen%
}{%
Strijbosch%
\ \protect \BOthers {.}}{%
{\protect \APACyear {2011}}%
}]{%
Strijbosch2011}
\APACinsertmetastar {%
Strijbosch2011}%
\begin{APACrefauthors}%
Strijbosch, L\BPBI W.%
, Syntetos, A\BPBI A.%
, Boylan, J\BPBI E.%
\BCBL {}\ \BBA {} Janssen, E.%
\end{APACrefauthors}%
\unskip\
\newblock
\APACrefYearMonthDay{2011}{}{}.
\newblock
{\BBOQ}\APACrefatitle {{On the interaction between forecasting and stock control: The case of non-stationary demand}} {{On the interaction between forecasting and stock control: The case of non-stationary demand}}.{\BBCQ}
\newblock
\APACjournalVolNumPages{International Journal of Production Economics}{133}{1}{470--480}.
\newblock
\begin{APACrefURL} \url{http://dx.doi.org/10.1016/j.ijpe.2009.10.032} \end{APACrefURL}
\newblock
\begin{APACrefDOI} \doi{10.1016/j.ijpe.2009.10.032} \end{APACrefDOI}
\PrintBackRefs{\CurrentBib}

\bibitem [\protect \citeauthoryear {%
Visentin%
, Prestwich%
, Rossi%
\BCBL {}\ \BBA {} Tarim%
}{%
Visentin%
\ \protect \BOthers {.}}{%
{\protect \APACyear {2021}}%
}]{%
Visentin2021}
\APACinsertmetastar {%
Visentin2021}%
\begin{APACrefauthors}%
Visentin, A.%
, Prestwich, S.%
, Rossi, R.%
\BCBL {}\ \BBA {} Tarim, S\BPBI A.%
\end{APACrefauthors}%
\unskip\
\newblock
\APACrefYearMonthDay{2021}{}{}.
\newblock
{\BBOQ}\APACrefatitle {{Computing optimal (R, s, S) policy parameters by a hybrid of branch-and-bound and stochastic dynamic programming}} {{Computing optimal (R, s, S) policy parameters by a hybrid of branch-and-bound and stochastic dynamic programming}}.{\BBCQ}
\newblock
\APACjournalVolNumPages{European Journal of Operational Research}{294}{1}{91--99}.
\newblock
\begin{APACrefURL} \url{https://doi.org/10.1016/j.ejor.2021.01.012} \end{APACrefURL}
\newblock
\begin{APACrefDOI} \doi{10.1016/j.ejor.2021.01.012} \end{APACrefDOI}
\PrintBackRefs{\CurrentBib}

\bibitem [\protect \citeauthoryear {%
Xiang%
, Rossi%
, Martin-Barragan%
\BCBL {}\ \BBA {} Tarim%
}{%
Xiang%
\ \protect \BOthers {.}}{%
{\protect \APACyear {2018}}%
}]{%
Xiang2018}
\APACinsertmetastar {%
Xiang2018}%
\begin{APACrefauthors}%
Xiang, M.%
, Rossi, R.%
, Martin-Barragan, B.%
\BCBL {}\ \BBA {} Tarim, S\BPBI A.%
\end{APACrefauthors}%
\unskip\
\newblock
\APACrefYearMonthDay{2018}{}{}.
\newblock
{\BBOQ}\APACrefatitle {{Computing non-stationary (s, S) policies using mixed integer linear programming}} {{Computing non-stationary (s, S) policies using mixed integer linear programming}}.{\BBCQ}
\newblock
\APACjournalVolNumPages{European Journal of Operational Research}{271}{2}{490--500}.
\newblock
\begin{APACrefURL} \url{https://doi.org/10.1016/j.ejor.2018.05.030} \end{APACrefURL}
\newblock
\begin{APACrefDOI} \doi{10.1016/j.ejor.2018.05.030} \end{APACrefDOI}
\PrintBackRefs{\CurrentBib}

\end{thebibliography}
